\documentclass[usenatbib,twocolumn]{aastex61}
\usepackage[table]{}
\usepackage{wasysym}

\accepted{to PASP}

\shorttitle{Discovery of KPS1-b}
\shortauthors{Burdanov et al.}
\begin{document}

\title{KPS-1b: the first transiting exoplanet discovered \\ using an amateur astronomer's wide-field CCD data}

\correspondingauthor{Artem Burdanov}
\email{artem.burdanov@uliege.be}

\author[0000-0001-9892-2406]{Artem Burdanov}
%\altaffiliation{Based on the work started at the Ural Federal University (Yekaterinburg, Russia)}
\affiliation{Space sciences, Technologies and Astrophysics Research (STAR) Institute, Universit{\'e} de Li{\`e}ge, All{\'e}e du 6 Ao{\^u}t 17, 4000 Li{\`e}ge, Belgium}

\author{Paul Benni}
\affiliation{Acton Sky Portal (Private Observatory), Acton, MA, USA}

\author[0000-0001-7573-1535]{Eugene Sokov}
\affiliation{Central Astronomical Observatory at Pulkovo of Russian Academy of Sciences, Pulkovskoje shosse d. 65, St. Petersburg, Russia, 196140}
\affiliation{Special Astrophysical Observatory, Russian Academy of Sciences, Nizhnij Arkhyz, Russia, 369167}

\author{Vadim Krushinsky}
\affiliation{Ural Federal University, ul. Mira d. 19, Yekaterinburg, Russia, 620002}

\author{Alexander Popov}
\affiliation{Ural Federal University, ul. Mira d. 19, Yekaterinburg, Russia, 620002}

\author[0000-0001-6108-4808]{Laetitia Delrez}
\affiliation{Astrophysics Group, Cavendish Laboratory, J.J. Thomson Avenue, Cambridge CB3 0HE, UK}

\author{Michael Gillon}
\affiliation{Space sciences, Technologies and Astrophysics Research (STAR) Institute, Universit{\'e} de Li{\`e}ge, All{\'e}e du 6 Ao{\^u}t 17, 4000 Li{\`e}ge, Belgium}

\author{Guillaume H{\'e}brard}
\affiliation{Institut d'Astrophysique de Paris, UMR 7095 CNRS, Universit{\'e} Pierre \& Marie Curie, 98bis boulevard Arago, 75014 Paris, France}
\affiliation{Observatoire de Haute-Provence, Universit{\'e} d'Aix-Marseille \& CNRS, 04870 Saint Michel l' Observatoire, France}

\author{Magali Deleuil}
\affiliation{Aix Marseille Universit\'e, CNRS, LAM (Laboratoire d'Astrophysique de Marseille) UMR 7326, 13388 Marseille, France}

\author{Paul A. Wilson}
\affiliation{Leiden Observatory, Leiden University, Postbus 9513, 2300 RA Leiden, The Netherlands}
\affiliation{Institut d'Astrophysique de Paris, UMR 7095 CNRS, Universit{\'e} Pierre \& Marie Curie, 98bis boulevard Arago, 75014 Paris, France}

\author{Olivier Demangeon}
\affiliation{Instituto de Astrof{\'\i}sica e Ci\^encias do Espa\c{c}o, Universidade do Porto, CAUP, Rua das Estrelas, PT4150-762 Porto, Portugal}

\author[0000-0002-4746-0181]{\"Ozg\"ur Ba\c{s}t\"urk}
\affiliation{Ankara University, Faculty of Science, Department of Astronomy and Space Science, TR-06100 Tandogan, Ankara, Turkey}

\author{Erika Pak\v{s}tiene}
\affiliation{Institute of Theoretical Physics and Astronomy, Vilnius University, Saul\.{e}tekio al. 3, Vilnius 10257, Lithuania}

\author{Iraida Sokova}
\affiliation{Central Astronomical Observatory at Pulkovo of Russian Academy of Sciences, Pulkovskoje shosse d. 65, St. Petersburg, Russia, 196140}

\author{Sergei A. Rusov}
\affiliation{Central Astronomical Observatory at Pulkovo of Russian Academy of Sciences, Pulkovskoje shosse d. 65, St. Petersburg, Russia, 196140}

\author{Vladimir V. Dyachenko}
\affiliation{Special Astrophysical Observatory, Russian Academy of Sciences, Nizhnij Arkhyz, Russia, 369167}

\author{Denis A. Rastegaev}
\affiliation{Special Astrophysical Observatory, Russian Academy of Sciences, Nizhnij Arkhyz, Russia, 369167}

\author{Anatoliy Beskakotov}
\affiliation{Special Astrophysical Observatory, Russian Academy of Sciences, Nizhnij Arkhyz, Russia, 369167}

\author{Alessandro Marchini}
\affiliation{Astronomical Observatory - DSFTA, University of Siena, Via Roma 56, 53100 Siena, Italy}

\author{Marc Bretton}
\affiliation{Baronnies Proven\c{c}ales Observatory, Hautes Alpes - Parc Naturel R\' egional des Baronnies Proven\c {c}ales, 05150 Moydans, France} 

\author{Stan Shadick}
\affiliation{Physics and Engineering Physics Department, University of Saskatchewan, Saskatoon, SK, Canada, S7N 5E2}

\author{Kirill Ivanov}
\affiliation{Irkutsk State University, ul. Karla Marxa d. 1, Irkutsk, Russia, 664003}
%\collaboration{(AAS Journals Data Scientists collaboration)}



\begin{abstract}
We report the discovery of the transiting hot Jupiter KPS-1b. This exoplanet orbits a V~=~13.0 K1-type main-sequence star every 1.7~days, has a mass of $1.090_{-0.087}^{+0.086}$~$M_{\mathrm{Jup}}$ and a radius of $1.03_{-0.12}^{+0.13}$~$R_{\mathrm{Jup}}$. The discovery was made by the prototype Kourovka Planet Search (KPS) project, which used wide-field CCD data gathered by an amateur astronomer using readily available and relatively affordable equipment. Here we describe the equipment and observing technique used for the discovery of KPS-1b, its characterization with spectroscopic observations by the SOPHIE spectrograph and with high-precision photometry obtained with 1-m class telescopes. We also outline the KPS project evolution into the Galactic Plane eXoplanet survey (GPX). The discovery of KPS-1b represents a new major step of the contribution of amateur astronomers to the burgeoning field of exoplanetology.
\end{abstract}




\section{Introduction} \label{sec:intro}
A major fraction of all exoplanets confirmed to date transit their host stars \citep{2011A&A...532A..79S,2013PASP..125..989A,2014PASP..126..827H}. The eclipsing configuration of these planets enables us to measure their radii and masses by combining prior pieces of knowledge on their host star properties with photometric and radial velocity measurements, and thus to constrain their bulk compositions \citep{Winn2010}. With sufficient signal to noise ratios (SNR), it is also possible to explore their atmospheric properties \citep{Winn2010,2014Natur.513..345B,2016SSRv..205..285M}. 

Despite their rarity in the Milky Way ($\sim$1\% occurrence rate, e.g. \citealt{2012ApJ...753..160W}), hot Jupiters (i.e. planets with minimum mass $M\,\sin i > 0.5~M_{\mathrm{Jup}}$ and orbital period $P~\lesssim~10$~days) are particularly favorable targets for non-space surveys. It is due to their short orbital periods that maximize their transit probabilities, and to their large sizes (up to 2 $R_{\mathrm{Jup}}$) that maximize their transit depths. These short period gas giants orbit in strong gravitational and magnetic fields \citep{2010exop.book..239C, 2010ApJ...708.1692C} and undergo irradiation larger than any planet in the Solar System \citep{2007ApJ...659.1661F}. Thorough characterization of hot Jupiters gives us an opportunity to research the response of these planets to severe environments, and to find out their planetary structures and chemical compositions. Moreover, their orbital obliquity can be measured during their transits \citep{2011ApJ...743...61S,2011ApJ...733..127S,2011A&A...527L..11H,2011A&A...533A.113M,2008A&A...488..763H}, bringing a key constraint on their dynamical history \citep{2010A&A...524A..25T}.

\begin{figure}
\centering
\includegraphics[width=1.05\columnwidth]{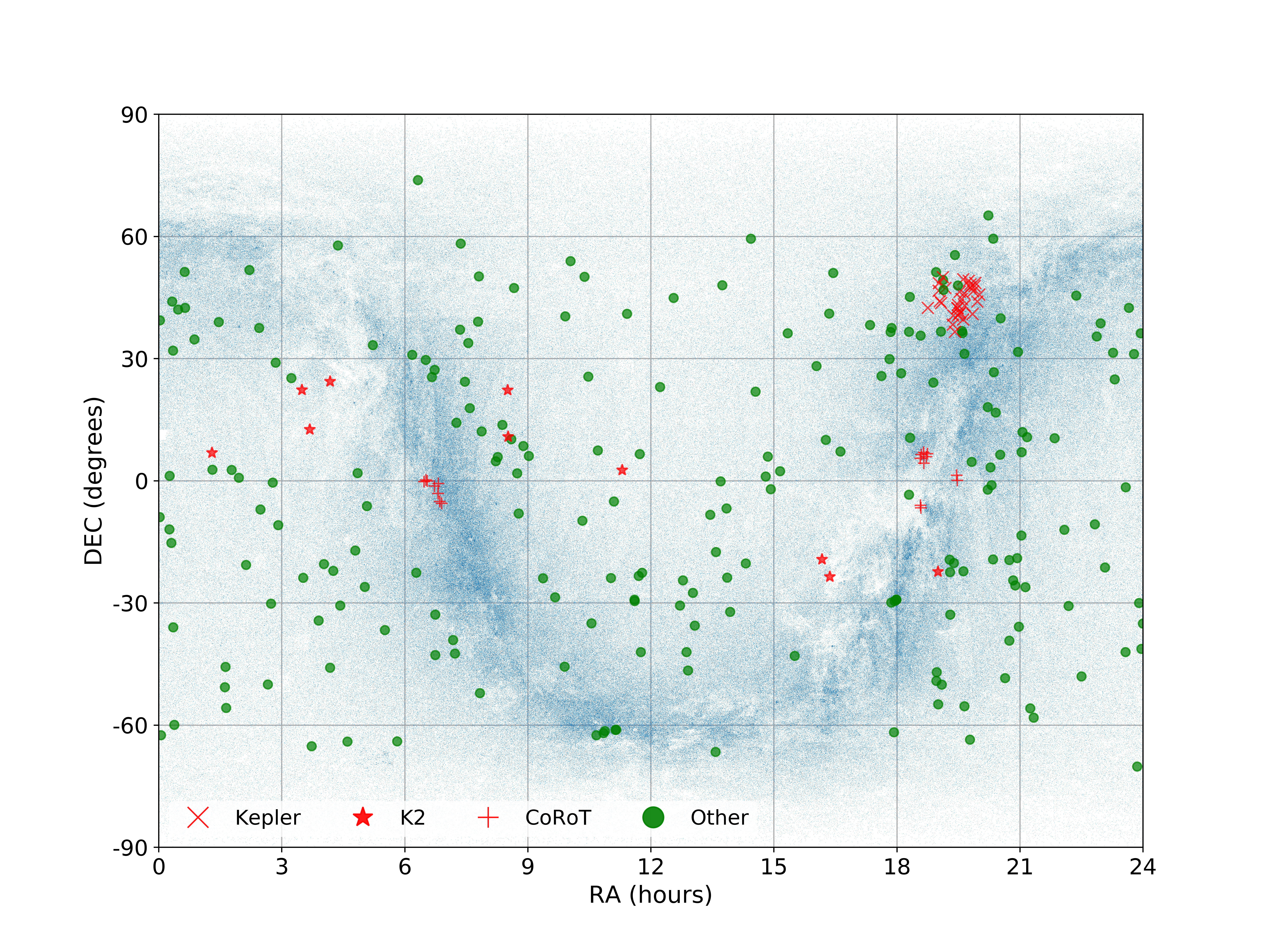}
\caption{Transiting hot Jupiters known to date on a projection of the celestial sphere. The exoplanet sample is from the NASA Exoplanet Archive \citep{2013PASP..125..989A} and stars' positions are from the Tycho-2 catalogue \citep{2000A&A...355L..27H}.}
\label{Fig_Known_HJ}
\end{figure}

$\sim$300 transiting hot Jupiters are known to date. Wide-field photometric surveys operating from the ground, such as WASP \citep{2006PASP..118.1407P}, HAT \citep{Bakos2004,2013PASP..125..154B}, KELT \citep{2012ApJ...761..123S}, Qatar \citep{Alsubai2013}, XO \citep{McCullough2005,2017arXiv171207275C} and TrES \citep{Alonso2004} discovered most of them. Space missions, namely \textit{CoRoT} \citep{Auvergne2009}, \textit{Kepler} \citep{Borucki2010} and \textit{K2} \citep{2014PASP..126..398H} also contributed to the number of known hot Jupiters. However, the combination of these surveys has left a significant fraction of the sky -- the Galactic plane (see Figure~\ref{Fig_Known_HJ}) -- relatively unexplored. Indeed, all these wide-field surveys, including future space missions like \textit{PLATO} \citep{2014ExA....38..249R} and \textit{TESS} \citep{2015JATIS...1a4003R}, have a poor spatial resolution in common. It translates into a large level of blending for the crowded fields of the Galactic plane, resulting into a poorer detection potential (due to signal dilution) combined with a drastically increased level of false-alarm probability (due to blended eclipsing binaries). For these reasons, most of the cited surveys try to avoid fields of low Galactic latitudes. Hence, many more hot Jupiters still remain undiscovered there.

This work is based on the hypothesis that amateur astronomy could play a significant role in their discovery. The synergy between amateur and professional astronomers in the exoplanetary research field has proven to be fruitful \citep{2012JAVSO..40..456C}. Some examples of that are  Zooniverse-based\footnote{\url{www.zooniverse.org}} projects Planet Hunters \citep{2012MNRAS.419.2900F} and Exoplanet Explorers\footnote{\url{www.zooniverse.org/projects/ianc2/exoplanet-explorers}}. Amateurs also participate in the ground-based exoplanet follow-up -- for example, KELT-FUN observations for KELT exoplanet candidates \citep{2017AAS...22914636M}, observations of KOI-1257 b \textit{Kepler} exoplanet candidate \citep{2014A&A...571A..37S}, and take an active part in search for Transit Timing Variations (TTVs, \citealt{2015MNRAS.450.3101B}). However, aforementioned projects are focused on searching for planets in already existing data sets, observing exoplanet candidates and already known planet-bearing stars discovered by professional astronomers. Now amateurs can also contribute to the search for transiting exoplanets by directly gathering wide-field CCD data. This approach is used by the PANOPTES \citep{2014SPIE.9145E..3VG} project and the project described in this paper. 

In 2012--2016 we carried out the Kourovka Planet Search (KPS, \citealt{2016MNRAS.461.3854B, 2015PZP....15....7P}) -- a prototype survey with the primary objective of searching for new transiting exoplanets in the fields of low Galactic latitudes. We used the MASTER-II-Ural robotic telescope \citep{Gorbovskoy2013} operated at the Kourovka astronomical observatory of the Ural Federal University in the Middle Urals (Russia). At first our observations of target fields TF1 and TF2 in the Galactic plane were performed solely with the MASTER-II-Ural telescope. But since 2014 the Rowe-Ackermann Schmidt Astrograph, operated at the private Acton Sky Portal Observatory (MA, USA), joined the project by observing an additional target field TF3 (located not in the Galactic plane, but in the Ursa Major constellation). Exoplanet candidates that were discovered in the fields TF1 and TF2 turned out to be astrophysical false positives. Here we announce the discovery of the transiting hot Jupiter KPS-1b, which was found after analyzing the TF3 wide-field CCD data gathered at the private Acton Sky Portal Observatory. 

The rest of this paper is structured as follows: information about initial detection observations and subsequent follow-up observations are given in Section \ref{sec:obs}; data analysis and derived parameters of planetary system are presented in Section~\ref{sec:params}; in Section \ref{sec:discuss} we briefly discuss the discovered planetary system, outline evolution of the KPS project and future prospects.

\section{Observations}\label{sec:obs}
\subsection{KPS transit detection photometry}
As a part of the KPS prototype survey, we observed target field \#3 (TF3), which is located in the Ursa Major constellation and contains the smallest number of stars of all the fields observed within the survey. This field is not in the plane of the Milky Way (poorly observable in the northern hemisphere at the time), but was nonetheless observed as part of the test observations with the Rowe-Ackermann Schmidt Astrograph (RASA) telescope for 21 nights in the period from January to April 2015 in the $R_c$ filter and with 50~sec exposures. The RASA telescope (Celestron Inc., Torrance, CA, USA) has a Schmidt-like optical design with a diameter of 279 mm and focal a length of 620 mm coupled with a Celestron CGEM mount. The telescope is based at the Acton Sky Portal private observatory in Acton, MA (USA). Software control of the telescope targeting and imaging was done by TheSkyX (Software Bisque, Golden, CO, USA). We used a SBIG ST-8300M camera (Santa Barbara Imaging Group, Santa Barbara, CA, USA) mounted on a custom adapter with a $R_c$ filter. One $1.67\times1.26$~deg$^2$ field was imaged with a pixel resolution of 1.79~arcsec~pixel$^{-1}$. Typical full width at half maximum (FWHM) of the stellar point spread function (PSF) values in the images were 2--3~arcsec.

\begin{figure}[h]
\centering
\includegraphics[width=0.47\textwidth]{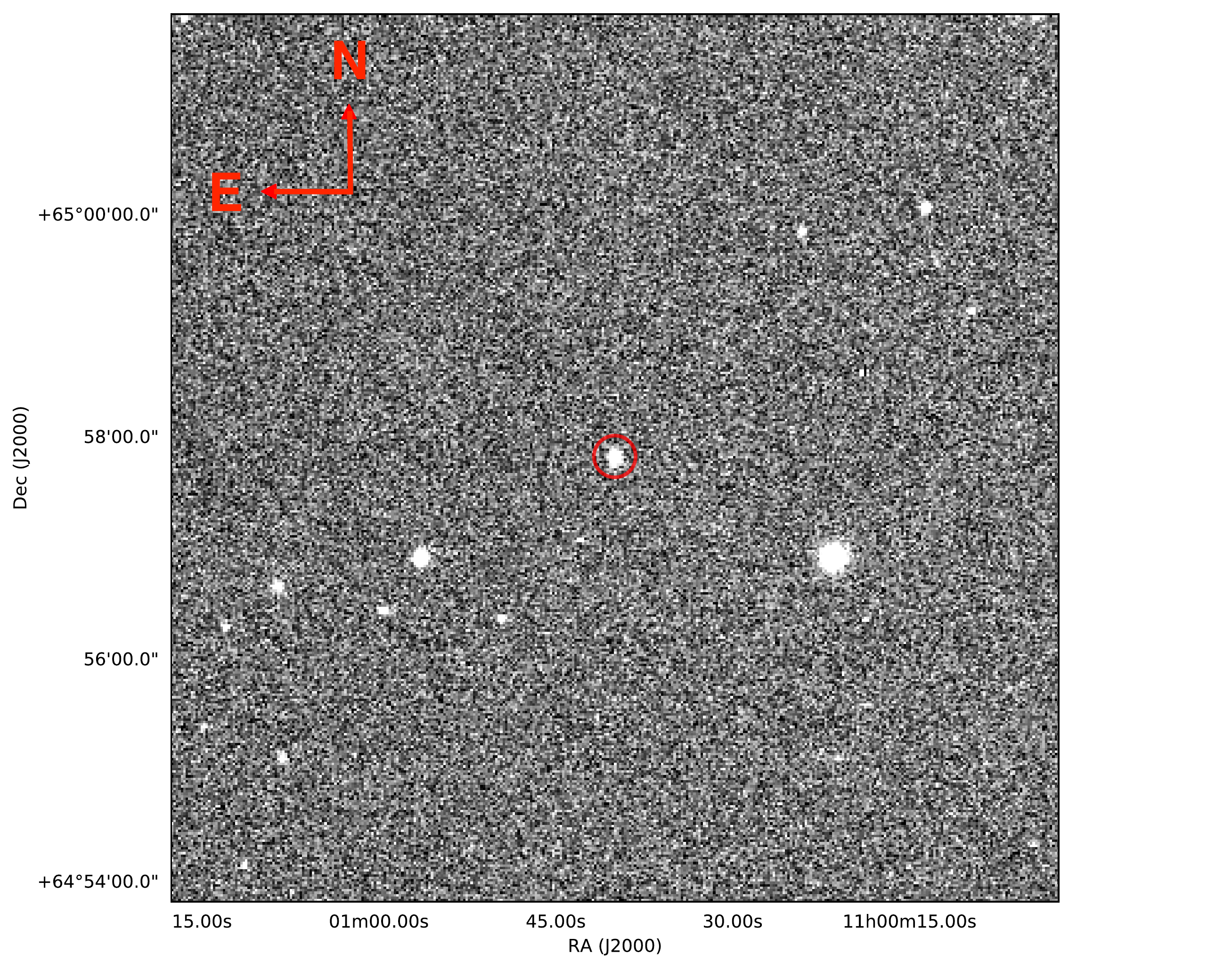}
\caption{Finding chart including KPS-1 host star as obtained with the MTM-500 telescope in V band.} 
\label{fig:finding_chart}
\end{figure}

\begin{figure}[h]
\includegraphics[width=\columnwidth]{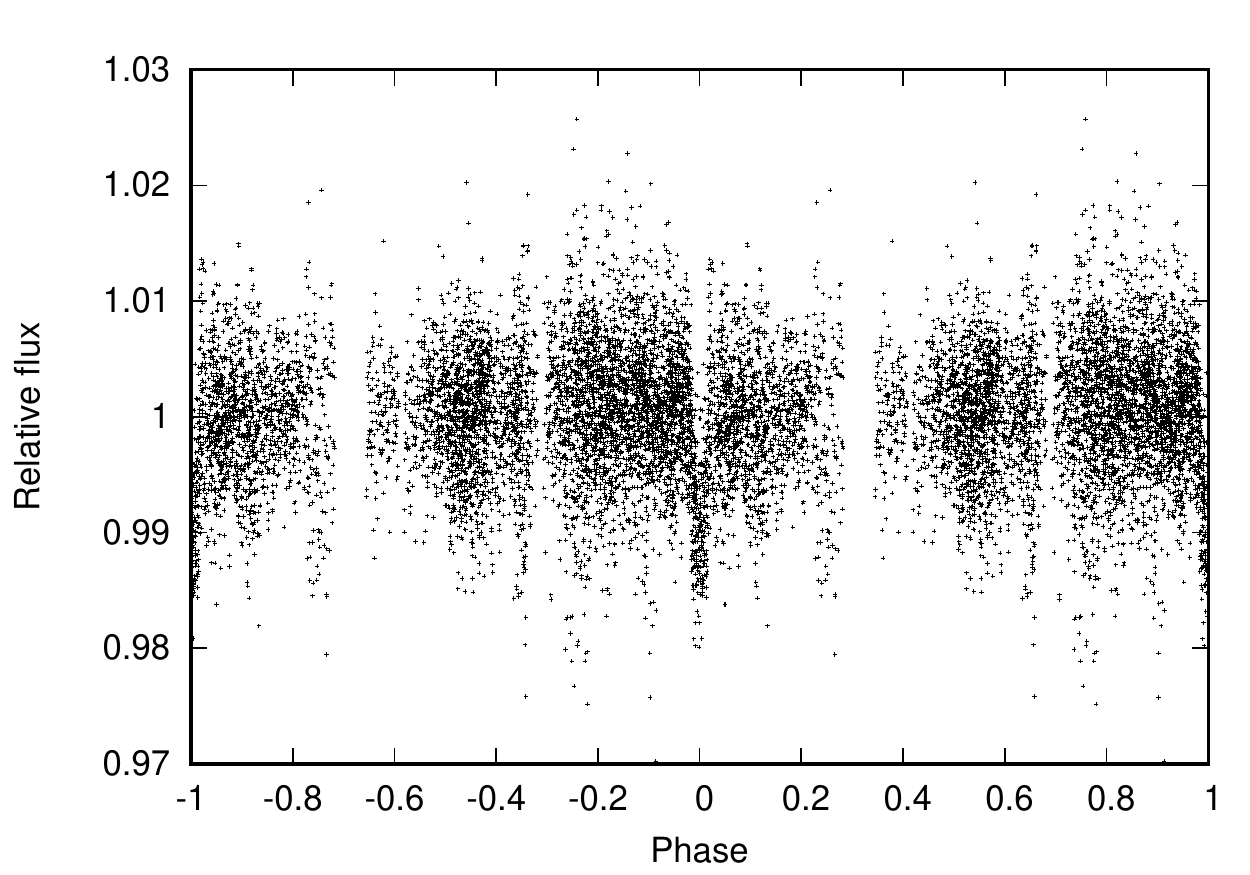}
\caption{Phase folded RASA telescope photometry for KPS-1b.} 
\label{fig_phase}
\end{figure}

We used Box-fitting Least Squares method (BLS, \citealt{2002A&A...391..369K}) to search for periodic transit-like signals in our time-series and found a significant peak in the periodogram of the star KPS-1 (KPS-TF3-663 = 2MASS 11004017+6457504 = GSC 4148-0138 = UCAC4775-030421, Figure~\ref{fig:finding_chart} and Table~\ref{tab:general_info}) located in the TF3 field. We detected five $\sim$~0.01~mag transit-like events, two of them were full, which allowed us etimate the period as $\sim$1.7~days. The phase folded light curve with this period is shown in Figure~\ref{fig_phase}. The details of data reduction, transiting signals search and research of exoplanet candidates can be found in \cite{2016MNRAS.461.3854B}.

\begin{table}[h]
\centering 
\caption{General information about KPS-1.}
\begin{tabular}{ll}
\hline
\hline
Identifiers & KPS-1\\
& KPS-TF3-663\tablenotemark{a}\\
& 2MASS11004017+6457504\\
& UCAC4775-030421\\
RA (J2000) & 11h 00m 40.176s\\ 
DEC (J2000) & +64d 57m 50.47s\\ 
Bmag\tablenotemark{b} & 13.949 \\
Vmag\tablenotemark{b} & 13.033\\
Jmag\tablenotemark{c} & 11.410\\
\hline
\end{tabular}

\tablenotetext{a}{Exoplanet candidate ID from \cite{2016MNRAS.461.3854B}}
\tablenotetext{b}{Data from the APASS catalog \citep{2016yCat.2336....0H}}
\tablenotetext{c}{Data from the 2MASS catalog \citep{2003yCat.2246....0C}}
\label{tab:general_info}
\end{table}

\subsection{Photometric follow-up observations}
We conducted a set of photometric follow-up runs to validate transit-like events visible in our wide field data, check for presence of the secondary minimum to reject possible eclipsing binary scenario and then to analyze the shape of the transit signal. For this purpose we used a network of telescopes established within the framework of the KPS survey (for a full list of telescopes see \citealt{2016MNRAS.461.3854B}).

% \begin{table}[h]
% \small
% \caption{Information about the telescopes that took part in photometric follow-up observations of KPS-1b.}
% \centering
% \begin{tabular}{cc}\hline
% \hline
% Telescope & Observatory\\\hline
% 0.28-m Celestron EdgeHD & Acton Sky Portal\\
% 11'' SCT & \\
% 0.5-m MTM-500 & Kislovodsk Mountain\\
% & Astronomical Station\\ 
% 1-m T100 & T\"{U}B\.{I}TAK National\\
% & Observatory\\
% 0.43-m Cassegrain & Baronnies Proven\c{c}ales\\
% & Observatory\\ 
% 0.3-m Smith-Cassegrain & Observatori Montcabrer\\
% Meade LX200 12'' and 14'' & University of Saskatchewan\\
% 0.3-m Maksutov-Cassegrain & University of Siena\\
% 1.65-m & Mol\.{e}tai Astronomical\\& Observatory \\
% \hline
% \end{tabular}
% \label{tab:telescopes}
% \end{table}

To determine the system's parameters, we used only high-precision photometric observations. These observations were conducted with the 1-m T100 telescope located at the T\"{U}B\.{I}TAK National Observatory (Turkey) and the 1.65-m telescope located at the Mol\.{e}tai Astronomical Observatory (Lithuania). 

\begin{figure}[h]
\includegraphics[width=\columnwidth]{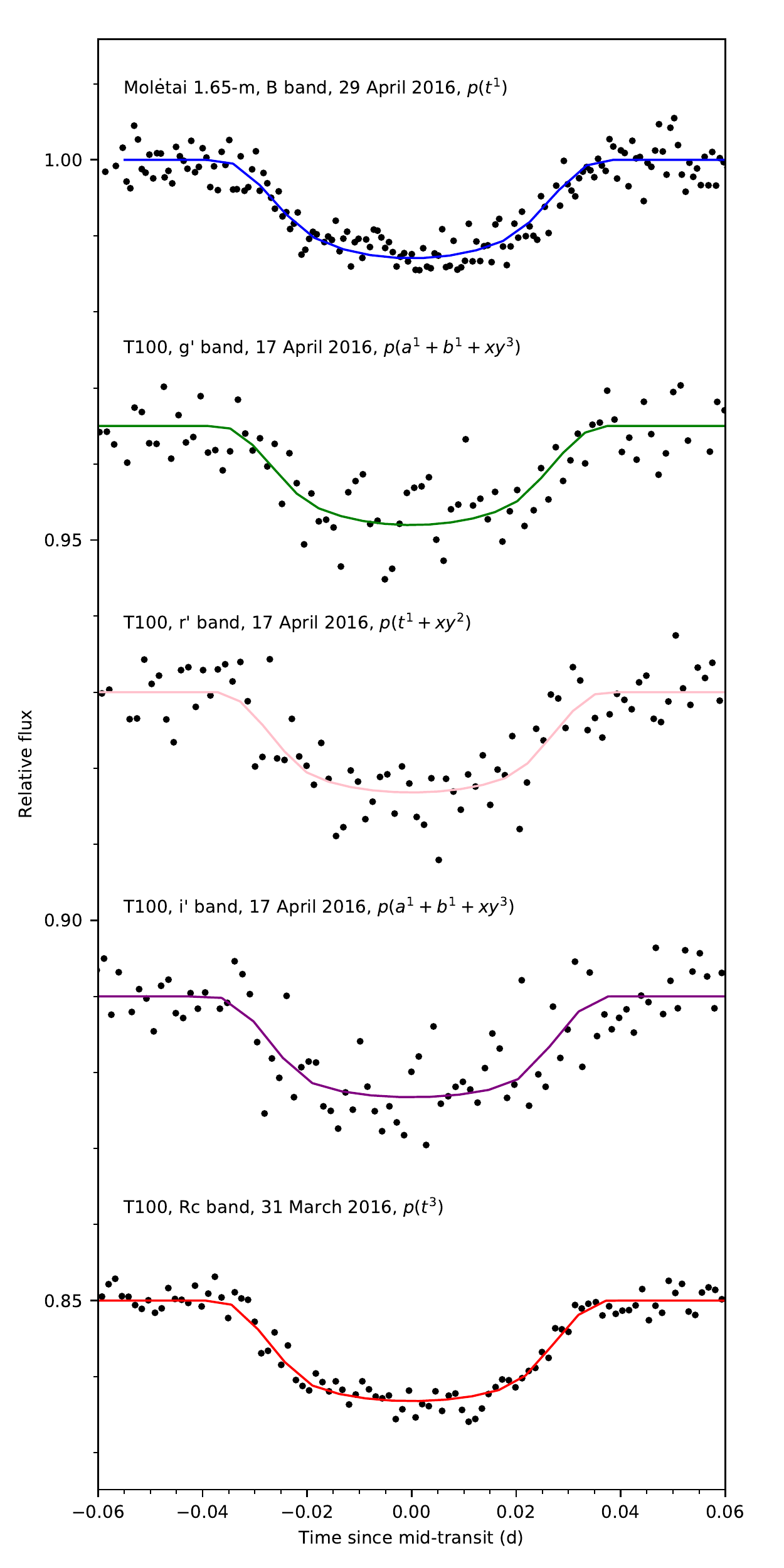}
\caption{Transits of KPS-1b in different bands as obtained with T100 telescope of T\"{U}B\.{I}TAK National Observatory of Turkey and 1.65-m Ritchey-Chr\'{e}tien telescope of the Mol\.{e}tai Astronomical Observatory (MAO, Lithuania). The observations are period-folded, multiplied by the baseline polynomials, and are shown in black. For each transit, the solid line represents the best-fit model. Along with the information about filter used, the baseline function is given (see section \ref{sec:global_mod} for details): $p(x^N)$  stands for a $N$-order polynomial function of airmass ($x=a$), positions of the stars on CCD ($x=xy$), time ($x=t$) and sky background ($x=b$).} 
\label{precise_lcs}
\end{figure}

T100 is the 1-m Ritchey-Chr\'{e}tien telescope at the T\"{U}B\.{I}TAK National Observatory of Turkey located at an altitude of 2500~m above the sea level. The telescope is equipped with a high quality Spectral Instruments 1100 CCD detector, which has 4096x4037 matrix of 15~$\mu$m-sized pixels. It was cooled down to -95~$^{\circ}$C with a Cryo-cooler. The pixel scale of the telescope is 0.31 arcsec~pixel$^{-1}$, that translates into 21.5$\times$21.5~arcmin$^2$ field of view (FOV). We used 2$\times$2 binning in order to reduce the read-out time to 15 seconds, which also resulted in a better timing resolution. The star image was slightly defocused to limit the systematic noise originating from pixel to pixel sensitivity variations. To compensate the defocussing, the exposure time was increased to 26, 13, and 14 seconds in the ASAHI\footnote{\url{www.asahi-spectra.com/opticalfilters/sdss.html}} SDSS g', r', and i' filters respectively in the transit observations of KPS-1b on 17 April 2016. A Johnson \textit{$R_c$} filter was used to observe a transit of KPS-1b on 31 March 2016 with 90 seconds exposure time. T100 has a very precise tracking system, improved even further by its auto-guiding system based on the short-exposure images acquired by an SBIG ST-402M camera. Thanks to this guiding system, the target was maintained within a few pixels during both observing runs. Data reduction consisted of standard calibration steps (bias, dark and flat-field corrections) and subsequent aperture photometry using IRAF/DAOPHOT \citep{Tody1986}. Comparison stars and aperture size were selected manually to ensure the best photometric quality in terms of the flux standard deviation of the check star 2MASS~J10595770+6455208, i.e. non-variable star similar in terms of magnitude and color to the target star.

Observations of KPS-1b transit on 29 April 2016 were conducted at the Mol\.{e}tai Astronomical Observatory (MAO, Lithuania) with the 1.65-m Ritchey-Chr\'{e}tien telescope. The telescope is equipped with the Apogee Alta U47-MB thermoelectrically cooled CCD camera, which has 1024x1024 matrix of 13~$\mu$m-sized  pixels. The pixel scale of the telescope is 0.51 arcsec~pixel$^{-1}$, what gives 8.7$\times$8.7~arcmin$^2$ FOV. Observations were conducted using a Johnson B filter with 60 seconds exposure time. The observed images were processed with the Muniwin program v.\,2.1.19 of the software package C-Munipack \citep{2014ascl.soft02006H}. Reduction also consisted of standard calibration steps and subsequent aperture photometry.

Obtained light curves show 1.3\%-depth achromatic transits. The light curves with imposed models and used baseline models (see section \ref{sec:global_mod} for details) are presented in Figure~\ref{precise_lcs}.

\begin{figure}[h]
\includegraphics[width=\columnwidth]{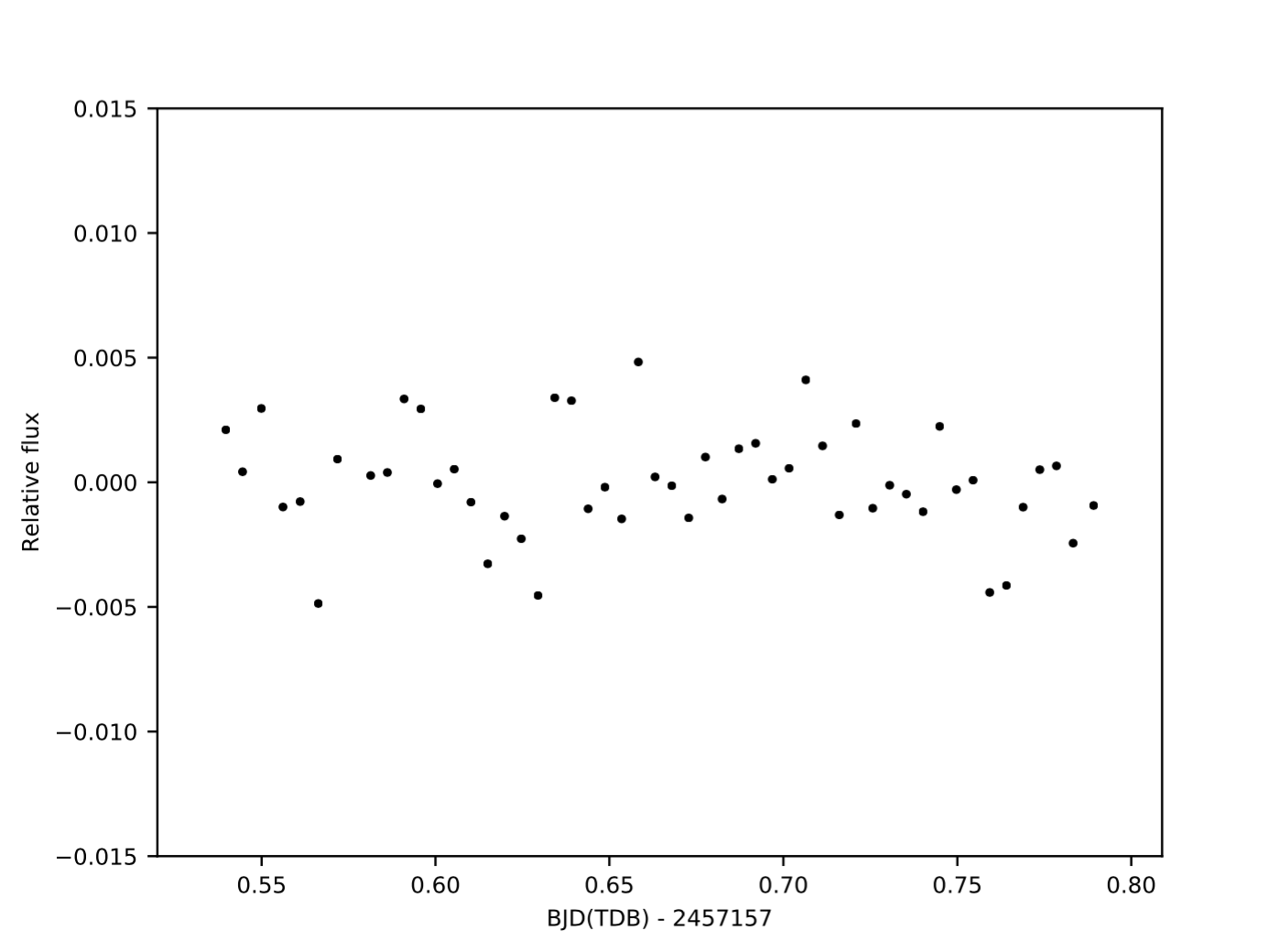}
\caption{Light curve of KPS-1 during the time of a possible secondary minimum (mid-eclipse time $BJD_{\mathrm{TDB}}=2457157.7284\pm0.0005$). Data was obtained with the 279-mm Celestron C11 telescope at the Acton Sky Portal (Acton, MA USA) in blue-blicking exoplanet filter and binned per 7~min intervals.} 
\label{secondary_minimum}
\end{figure}

As part of the KPS exoplanet candidates vetting, we perform photometric observations during the times of possible secondary minimums. Visual inspections of such light curves help to reveal obvious eclipsing binary nature of some of the candidates (for example, KPS-TF1-3154 and KPS-TF2-11789 from \citealt{ 2016MNRAS.461.3854B}). We observed KPS-1 at the expected time of secondary minimum on 15 May 2015. We used the 279-mm Celestron C11 telescope at the Acton Sky Portal. The telescope is equipped with the SBIG ST8-XME (Diffraction Limited / SBIG, Ottawa, Canada) thermoelectrically cooled CCD camera, which has 1530x1020 matrix of 9~$\mu$m-sized  pixels. FOV is 24.2$\times$16.2~arcmin$^2$ with the pixel scale of 0.95 arcsec~pixel$^{-1}$. Observations were conducted using an Astrodon Blue-blocking Exo-Planet filter\footnote{\url{www.astrodon.com/store/p9/Exo-Planet_Filter.html}} which has a transmittance over 90\% from 500~nm to beyond 1000~nm with 40 seconds exposure times. The observed images were processed in the similar way as the images from T100 and 1.65-m telescope, but with the AIP4WIN2 program (Willmann-Bell, Inc. Richmond, VA USA). These observations produced no detection (see Figure~\ref{secondary_minimum}) and allowed us to rule out any eclipsing binary scenario with secondary eclipses deeper than 1\% (3-sigma upper limit), thus strengthening the planetary hypothesis.

%\startlongtable
%\begin{deluxetable*}{c|c|c|c|c|c}
%\tablecaption{Summary of the follow-up transit photometry of KPS-1b %\label{tab:table}}
%\tablehead{
%\colhead{Date (UT)} & \colhead{Telescope} & \colhead{Observatory} & \colhead{Filter}  & \colhead{$T_{exp}$}  & \colhead{$N_{p}$} \\
%\startdata
%2015-04-28 & MTM500 & Kislovodsk Mountain \\ Astronomical Station & Clear & 
%\enddata
%}
%\colnumbers
%\startdata
%\enddata
%\end{deluxetable*}
%\endlongtable

\subsection{Speckle-interferometric observations}
\begin{figure}[h]
\includegraphics[width=\columnwidth]{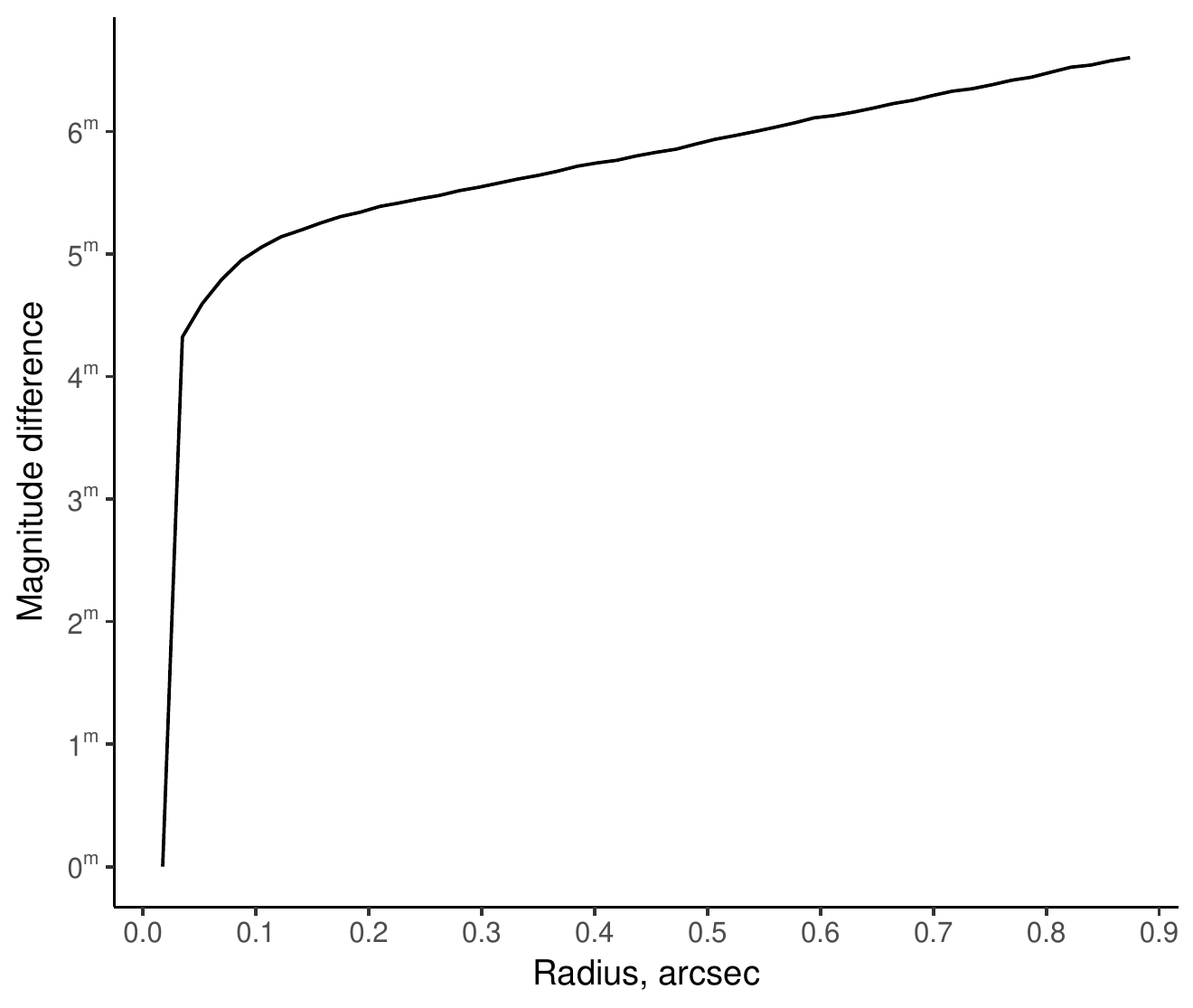}
\caption{Auto-correlation function derived from the power spectrum by Fourier transform of KPS-1 star are calculated, translated to the $3\sigma$ detection limit $\Delta m_3$ and plotted as a function of radius.} 
\label{KPS_limits_bw}
\end{figure}

On 7 June 2015, we conducted speckle-interferometric observations of KPS-1 host-star with the 6-m telescope of the SAO RAS. We used a speckle interferometer based on EMCCD \citep{2009AstBu..64..296M} with a new Andor iXon Ultra DU-897-CS0 detector. Observations were carried out in a field of $4.5 \times 4.5$~arcsec$^2$ in 800~nm filter with a bandwidth of 100~nm. We obtained two series of speckle images with exposure times of 20 and 100 milliseconds. During the analysis of power spectra we did not find any components at an angular distance up to 0.04 arcsec with a brightness difference of up to 3 magnitudes and at a distance of up to 0.06 arcsec with a brightness difference of up to 4 magnitudes (see Figure~\ref{KPS_limits_bw}).

\subsection{Spectroscopic observations}\label{subsec:spec_obs}

We carried out spectroscopic follow-up observations of KPS-1 with the SOPHIE high-resolution spectrograph mounted on the 1.93-m telescope at the Haute-Provence Observatory in south-eastern France \citep{2009A&A...505..853B}. This instrument provides high-precision radial velocity measurements, which allowed us to determine the planetary nature of the event and then to characterize the secured planet by measuring its mass and constraining its eccentricity. 

Observations of KPS-1 with SOPHIE were conducted in the High-Efficiency mode (resolving power $R=40\,000$). Ten spectra were secured between March and July 2016, with exposure times ranging from 1150 to 2340 seconds depending on the weather conditions. The SOPHIE automatic pipeline was used to extract the spectra. SNR per pixel at 550~nm were between 13 and 25 depending on the exposures, with typical values around 23. The weighted cross-correlation function (CCF) with a G2-type numerical mask was used to measure the radial velocities \citep{1996A&AS..119..373B,2002A&A...388..632P}. Due to poor SNRs, we removed 12 bluer of the 38 spectral orders from the CCF; this improved the accuracy of the radial-velocity measurement but did not significantly change the result by comparison to CCFs made on the whole SOPHIE spectral range.

We computed the error bars on radial velocities and the CCF bisectors following \cite{2010A&A...523A..88B}. Moonlight affected six spectra, but this contamination was corrected using the second fiber aperture pointing the sky during the observations \citep{2008MNRAS.385.1576P,2008A&A...488..763H}. It allowed us to introduce large radial velocity corrections from 100 up to 650~m/s.

\begin{figure}
\includegraphics[width=1.0\columnwidth]{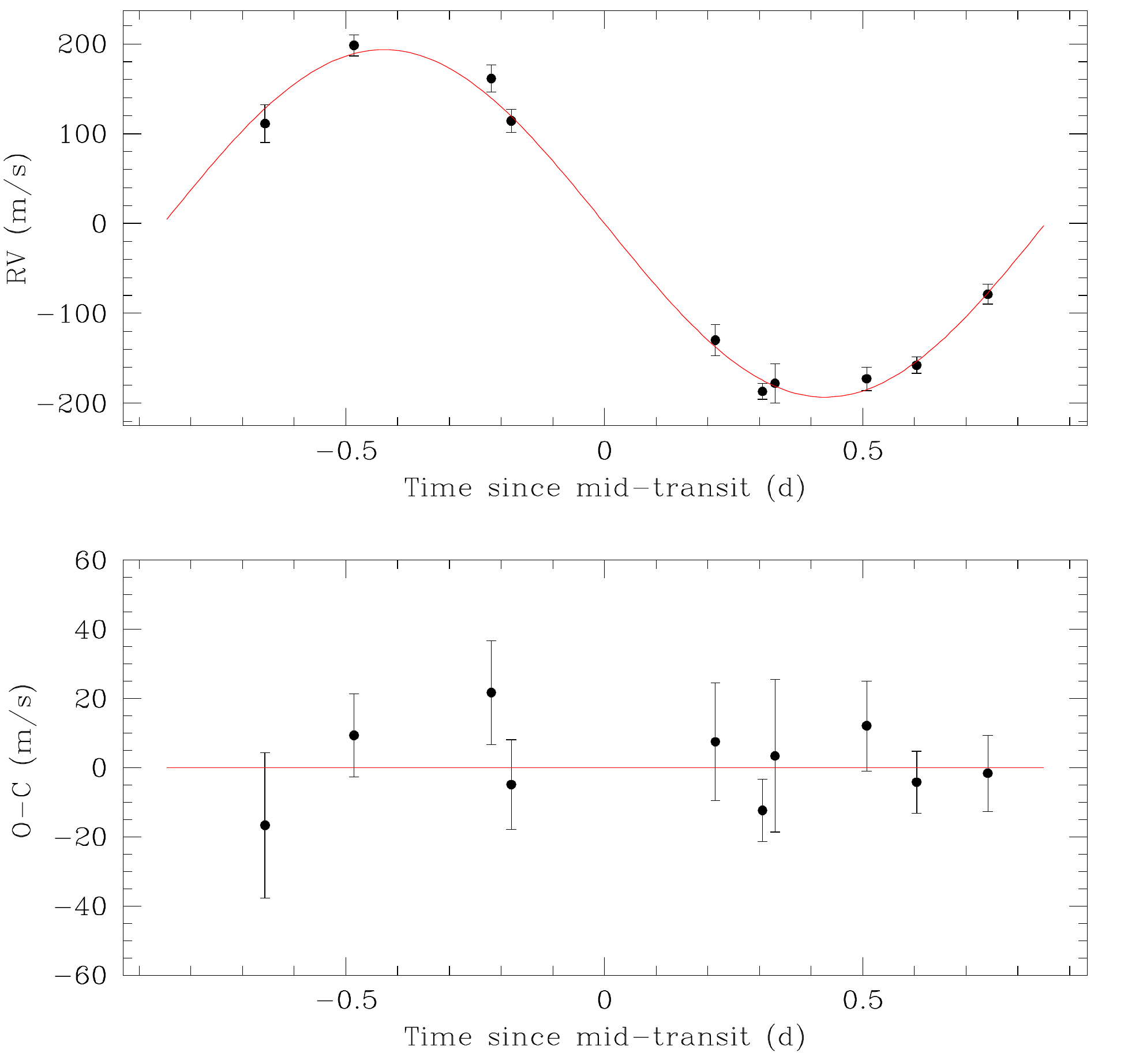}
\caption{SOPHIE radial velocity measurements of KPS-1 phase folded at the 1.7~days period of the exoplanet with the imposed best-fit Keplerian model.} 
\label{fig:RVs}
\end{figure}

The radial velocity measurements and the CCF bisectors are provided in Table~\ref{tab:RVs} and are shown in Figure~\ref{fig:RVs} with the imposed Keplerian fits and the residuals. 

\begin{table}[h]
\small
\caption{SOPHIE radial velocities measurements (RV) for KPS-1.}
\centering
\begin{tabular}{cccc}\hline
\hline
Epoch & RV & $\sigma_{RV}$ & Bisector\\
 (BJD - 2 450 000) & (km/s) & (km/s) & (km/s)\\\hline
7465.5366 &   -15.722  &       0.013 &          -0.057\\
7466.4585 &   -15.915  &       0.011 &          -0.060\\
7468.4725 &   -15.725  &       0.021 &          -0.015\\
7515.5258 &   -16.014  &       0.022 &          -0.009\\
7539.3881 &   -16.023  &       0.009 &          -0.013\\
7541.3926 &   -15.994  &       0.009 &          -0.024\\
7576.3995 &   -15.675  &       0.015 &          -0.048\\
7585.3637 &   -15.966  &       0.017 &          -0.067\\
7586.3709 &   -15.638  &       0.012 &          -0.027\\
7587.3628 &   -16.009  &       0.013 &          -0.006\\
\hline
\end{tabular}
\label{tab:RVs}
\end{table}

\begin{figure}
\includegraphics[width=1.0\columnwidth]{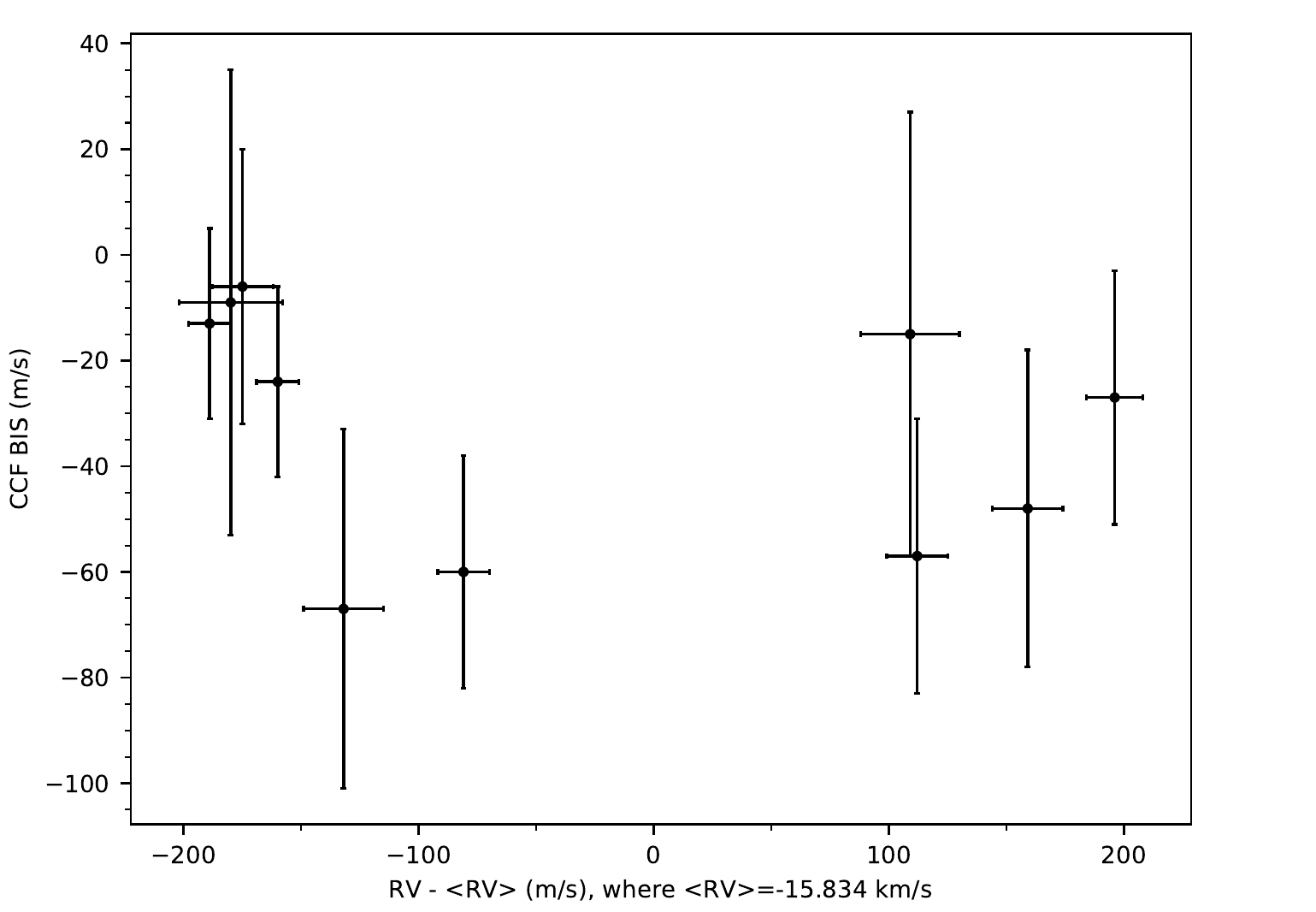}
\caption{CCF bisector spans as a function of measured radial velocities (RV).} 
\label{fig:ccf_bis}
\end{figure}

The radial velocity measurements have a semiamplitude around $K\sim200$\ m/s. Their variations are in phase with the transit ephemeris obtained from photometry, implying a Jupiter mass for the companion. We used different stellar masks (F0 or K5) to measure radial velocities and obtained similar amplitudes to those obtained with the G2 mask. We found no indication that these variations are caused by blended stars of different spectral types. Also there are no trends of the bisector spans as a function of the measured radial velocities (see~Figure~\ref{fig:ccf_bis}), what strengthens our conclusion that KPS-1 harbors a giant transiting exoplanet and velocity variations are not produced neither by stellar activity nor blended stars.

\section{Analysis}\label{sec:params}
\subsection{Stellar parameters from spectroscopy}

SOPHIE spectra were used to obtain the stellar parameters of KPS-1. Spectra were corrected for the radial velocities, then set in the rest frame and co-added. We analyzed isolated spectral lines in the Fourier space to deduce $v \sin i$ and the macroscopic velocity $v_{mac}$. Fitting of the observational spectra to the synthetic one was done using the SME package \citep{1996A&AS..118..595V}. Through the iterative process, we obtained the metallicity \lbrack Fe/H\rbrack, the stellar effective temperature $T_{\mathrm{eff}}$ using the hydrogen lines, log$\,g_{\star}$ using the Ca\,I and Mg\,I lines. Obtained results are presented in Table~\ref{tab:spectral_an}. 

\begin{table}[h]
\centering 
\caption{Stellar parameters from spectroscopy.}
\begin{tabular}{ll}
\hline
\hline
Spectral type & K1\\
$T_{\mathrm{eff}}$ & 5165$\pm$90~K\\
log$\,g_{\star}$ & 4.25$\pm$0.12~cgs\\
\lbrack Fe/H\rbrack & 0.22$\pm$0.13\\
$v_{mac}$ & $\sim1$~km/s\\
$v \sin i$ & 5.1$\pm$1.0~km/s\\
\hline
\end{tabular}
\label{tab:spectral_an}
\end{table}

\subsection{Global modeling of the data}\label{sec:global_mod}
We conducted a combined analysis of our radial velocity data obtained with a G2-type numerical mask (see sub-section \ref{subsec:spec_obs}) and available high-precision photometric light curves to put constraints on the KPS-1 system parameters. We used the Markov Chain Monte-Carlo (MCMC) method implemented in the code described in \citealt{2012A&A...542A...4G}. The code uses a Keplerian model by \citealt{2010exop.book...15M} and a transiting light curve model by \citealt{2002ApJ...580L.171M} to simultaneously treat all the available data.

Each light curve was multiplied by a different baseline model -- a polynomial with respect to the external parameters of the time-series (time, airmass, mean FWHM, background values, position of the stars on the CCD or combinations of these parameters). These models help to account for photometric variations related to external instrumental or environmental effects. We used minimization of the Bayesian Information Criterion (BIC, \citealt{1978AnSta...6..461S}) to select proper baseline models. Used baselines models can be found in Figure~\ref{precise_lcs}.

We used this set of parameters, which were perturbed randomly at each step of the chains (jump parameters): 
\begin{itemize}
  \item the orbital period $P$;
  \item the transit width (from first to last contact) $W_{14}$;
  \item the mid-transit time $T_0$;
  \item the ratio of the planet and star areas $(R_p/R_\star)^2$, where the planetary radius is $R_p$ and the stellar radius is $R_\star$; 
  \item the impact parameter $b^\prime = a\,cos\,i_p/R_\star$ (if orbital eccentricity $e=0$), where $a$ is the semi-major axis and $i_p$ is the orbital inclination; 
  \item the parameter $K_2 = K\sqrt{1 - e^2} P^{1/3}$, where $K$ is the radial velocity orbital semi-amplitude. $K_2$ is used instead of $K$ to minimize the correlation of the obtained uncertainties; 
  \item  $\sqrt{e}\,\sin \omega$ and $\sqrt{e}\,\cos \omega$ parameters, where $\omega$ is the argument of periastron;
  \item the effective temperature $T_{\mathrm{eff}}$ of the host star and its metallicity \lbrack Fe/H\rbrack;
  \item the combinations $c_1 =2u_1 +u_2$ and $c_2 = u_1 - 2u_2$ of the quadratic limb-darkening coefficients $u_1$ and $u_2$. To minimize the correlation of the obtained uncertainties, the combinations of limb-darkening coefficients $u_1$ and $u_2$ were used \citep{2006ApJ...652.1715H}.
\end{itemize}
Uniform non-informative prior distributions were assumed for the jump parameters for which we have no prior constraints.

%Prior distributions used for $u_1$ and $u_2$ are presented in Table \ref{tab:limb-dark}.
%\begin{table}
%\small
%\caption{Expectations and standard deviations of the normal distributions used as prior distributions for the quadratic limb-darkening (LD) coefficients $u_1$ and $u_2$ used in this work.}
%\label{tab:limb-dark}
%\centering
%\begin{tabular}{cc}\hline
%LD coefficient & KPS-1\\\hline
%$u_{1,B}$ & 0.841~$\pm$~0.033\\
%$u_{2,B}$ & -0.005~$\pm$~0.025\\
%$u_{1,R}$ & 0.563~$\pm$~0.033\\
%$u_{2,R}$ & -0.210~$\pm$~0.016\\
%$u_{1,g'}$ & 0.743~$\pm$~0.015\\
%$u_{2,g'}$ & 0.053~$\pm$~0.024\\
%$u_{1,r'}$ & 0.538~$\pm$~0.025\\
%$u_{2,r'}$ & 0.190~$\pm$~0.019\\
%$u_{1,i'}$ & 0.436~$\pm$~0.029\\
%$u_{2,i'}$ & 0.216~$\pm$~0.013\\
%\hline
%\end{tabular}
%\end{table}

Before the main analysis, we converted existing timings in $HJD_{\mathrm{UTC}}$ standard to the more practical $BJD_{\mathrm{TDB}}$ (see \citealt{2010PASP..122..935E} for details) and multiplied initial photometric errors by the correction factor $CF = \beta_{w} \times \beta_{r}$. Factor $\beta_{w}$ allows proper estimation of the white noise in our photometric time-series. It equals to the ratio of residuals standard deviation and mean photometric error. Factor $\beta_{r}$ allows us to consider red noise in our time-series and equals to the largest value of the standard deviation of the binned and unbinned residuals for different bin sizes. 

Five Markov chains with a length of 100 000 steps and with 20\% burn-in phase were performed. Each MCMC chain was started with randomly perturbed versions of the best BLS solution. We checked their convergences with using the statistical test of \cite{1992StaSc...7..457G} which was less than 1.11 for each jump parameter. Jump parameters and Kepler's third law were used to calculate stellar density $\rho_\star$ at each chain step (see e.g. \citealt{2003ApJ...585.1038S}, \citealt{Winn2010}). Then we calculated the stellar mass $M_{\star}$ using $T_{\mathrm{eff}}$ and [Fe/H] from spectroscopy and an observational law $M_{\star}$($\rho_{\star}$,$T_{\mathrm{eff}}$,[Fe/H]) \citep{2010A&A...516A..33E,2011A&A...533A..88G}. A sample of detached binary systems by \cite{2011MNRAS.417.2166S} was used to calibrate the empirical law. Then we derived the stellar radius $R_\star$ using $\rho_\star$ and $M_{\star}$. After that we deduced planet parameters from the set of jump parameters, $M_{\star}$ and $R_\star$.

We preformed two analyses: first, where eccentricity was allowed to float and second, where the orbit was circular. We adopt the results of the circular assumption as our nominal solutions and place a 3$\sigma$ upper limit on orbital eccentricity of 0.09. Derived parameters of the KPS-1 system are shown in Table~\ref{tab:sys_params}.

\begin{table*}
\centering 
\caption{Parameters of KPS-1 system.}
\begin{tabular}{lll}
\hline
\hline
\multicolumn{3}{c}{Parameters from global MCMC analysis}\\
\hline
Jump parameters & $e=0$ \textbf{(adopted)} & $e\geq0$\\
\hline
Orbital period $P$ [d] & $1.706291_{-0.000059}^{+0.000059}$ & $1.706290_{-0.000059}^{+0.000062}$\\
Transit width $W_{14}$ [d] & $0.0700_{-0.0020}^{+0.0023}$ & $0.0698_{-0.0019}^{+0.0022}$\\
Mid-transit time $T_0$ \lbrack $BJD_{\mathrm{TDB}}$\rbrack & $2457508.37019_{-0.00078}^{+0.00079}$ & $2457508.37019_{-0.00077}^{+0.00082}$\\
$b^\prime = a\,cos\,i_p/R_\star$ [$R_\star$] & $0.754_{-0.049}^{+0.040}$ & $0.759_{-0.050}^{+0.043}$\\
Planet/star area ratio $(R_p/R_\star)^2$ [\%] & $1.307_{-0.076}^{+0.085}$ & $1.300_{-0.072}^{+0.080}$\\
$K_2$ [m/s] & $237.4\pm6.5$ & $236.6\pm6.5$\\
$\sqrt{e}\,\sin \omega$ & 0 (fixed) & $0.07_{-0.12}^{+0.11}$\\
$\sqrt{e}\,\cos \omega$ & 0 (fixed) & $0.040_{-0.087}^{+0.075}$\\
%$C_{1,B}$ & \\
%$C_{2,B}$ & \\
%$C_{1,g'}$ & \\
%$C_{2,g'}$ & \\
%$C_{1,r'}$ & \\
%$C_{2,r'}$ & \\
%$C_{1,i'}$ & \\
%$C_{2,i'}$ & \\
$T_{\mathrm{eff}}$\tablenotemark{a} [K] & $5165\pm90$ & $5165\pm90$\\
Fe/H\tablenotemark{a}[dex] & $0.22\pm0.13$ & $0.22\pm0.13$\\
\hline
Deduced stellar parameters\\
\hline
Mass $M_{\star}$ [$M_{\astrosun}$] & $0.892_{-0.1}^{+0.09}$ & $0.894_{-0.09}^{+0.09}$\\
Radius $R_{\star}$ [$R_{\astrosun}$] & $0.907_{-0.082}^{+0.086}$ & $0.914_{-0.081}^{+0.086}$\\
Mean density $\rho_{\star}$ [$\rho_{\astrosun}$] & $1.19_{-0.23}^{+0.29}$ & $1.17_{-0.23}^{+0.28}$\\
Surface gravity log$\,g_{\star}$ [cgs] & $4.47\pm0.06$ & $4.47\pm0.06$\\
%$u_{1,B}$ & \\
%$u_{2,B}$ & \\
%$u_{1,g'}$ & \\
%$u_{2,g'}$ & \\
%$u_{1,r'}$ & \\
%$u_{2,r'}$ & \\
%$u_{1,i'}$ & \\
%$u_{2,i'}$ & \\
\hline
Deduced planet parameters\\
\hline
$K$ [m/s] & $198.7\pm5.5$ & $198.1\pm5.5$\\
Planet/star radius ratio $R_p/R_\star$ & $0.1143_{-0.0034}^{+0.0037}$ & $0.1140_{-0.0032}^{+0.0034}$\\
$b$ [$R_\star$] & $0.754_{-0.049}^{+0.040}$ & $0.751_{-0.049}^{+0.038}$\\
Orbital semi-major axis $a$ [au] & $0.0269\pm0.0010$ & $0.0269\pm0.0010$\\
Orbital eccentricity $e$ & 0 (fixed) & $0.017_{-0.012}^{+0.022} < 0.09$\\
Orbital inclination $i_p$ [deg] & $83.20_{-0.90}^{+0.88}$ & $83.12_{-0.93}^{+0.90}$\\
Argument of periastron $\omega$ [deg] & - & $60.470_{-0.009}^{+0.006}$\\
Surface gravity log$\,g_{p}$ [cgs] & $3.42\pm0.09$ & $3.42\pm0.09$\\
Mean density $\rho_{p}$ [$\rho_\mathrm{Jup}$] & $0.99_{-0.27}^{+0.37}$ & $0.98_{-0.26}^{+0.36}$\\
Mass $M_p$ [$M_\mathrm{Jup}$] & $1.090_{-0.087}^{+0.086}$ & $1.089_{-0.086}^{+0.085}$\\
Radius $R_p$ [$R_\mathrm{Jup}$] & $1.03_{-0.12}^{+0.13}$ & $1.04_{-0.11}^{+0.13}$\\
Roche limit $a_R$ [au] & $0.0113_{-0.0014}^{+0.0015}$ & $0.0113_{-0.0013}^{+0.0015}$\\
$a/a_{R}$ & $2.37\pm0.25$ & $2.37\pm0.25$\\
Equilibrium temperature $T_{eq}$ [K] & $1459\pm56$ & $1451\pm58$\\
Irradiation $I_p$ [$I_{\mathrm{Earth}}$] & $729_{-110}^{+130}$ & $736_{-110}^{+130}$\\
\hline
\end{tabular}
\tablenotetext{a}{Used as priors from the spectroscopic analysis}
\label{tab:sys_params}
\end{table*}

\section{Discussion and conclusions}\label{sec:discuss}

We presented a new transiting hot Jupiter KPS-1b discovered by the prototype Kourovka Planet Search (KPS) project, which used wide-field CCD data gathered by an amateur astronomer. KPS-1b is a typical hot Jupiter with a mass of $1.090_{-0.087}^{+0.086}$~$M_{\mathrm{Jup}}$ and a radius of $1.03_{-0.12}^{+0.13}$~$R_\mathrm{Jup}$ resulting in a mean density of $0.98_{-0.34}^{+0.27}$~$\rho_\mathrm{Jup}$. However, a period of 1.7 days is relatively short comparing to the majority of other known hot Jupiters \citep{2014PASP..126..827H}.

After this very promising discovery, we upgraded the original RASA telescope setup. A FLI ML16200 CCD camera (Finger Lakes Instrumentation, Lima, NY, USA) was installed. It has an area approximately twice that of the SBIG ST-8300M. The RASA telescope was moved to a Losmandy G11 mount (Hollywood General Machining Inc., Burbank, CA, USA). To increase sky coverage, four fields are imaged now (FOV: $2.52\times2.02$~deg$^2$ per field) on a cycle basis to obtain an overall FOV of $8.0\times2.5$~deg$^2$. The adjacent fields keep a constant right ascension while the declination is changed, which prevented problems with the equatorial mount when meridian flipping from east to west. The image scale increased from 1.8 to 2.02 arcsec~pixel$^{-1}$. 

Instead of autoguiding, the telescope was resynchronized before imaging field \#1 with a plate solve image.  CCD Commander\footnote{Matt Thomas, \url{www.ccdcommander.com}} control software scripts working with TheSkyX was used for multi-field targeting and plate solve resynching. Three images with 40 second exposures are taken per field to allow for a 3-point median filter to be applied on the processed photometry light curves to reduce noise. The overall cadence of the four-field cycle is about 10-11 minutes.

From what had been learned from the KPS survey, the Acton Sky Portal based exoplanet survey telescope now preferably targets the high star density fields of the Milky Way to increase the odds of discovering stars with hot Jupiter exoplanets. This new survey is called the Galactic Plane eXoplanet Survey (GPX). The goal of GPX is to record $\sim$150 hours of data over a three-month period while the targeted high-density star fields were observable. This is possible because of the high 2.02 arcsec~pixel$^{-1}$ resolution where other low resolution surveys (13.7-23 arcsec~pixel$^{-1}$ in case of HAT, WASP, KELT and XO) would encounter challenges because of the blending false-positives scenarios. Similar in terms of the pixel scale (5~arcsec~pixel$^{-1}$), the NGTS survey observes fields which are mostly more than 20~deg from the Galactic plane \citep{2017arXiv171011100W}.

The K-pipe data processing pipeline \citep{2016MNRAS.461.3854B} was also improved for GPX. With the three images per field data, a 3-point median filter of the post photometry processed star light curves was added to Astrokit \citep{2014AstBu..69..368B} to reduce photometric noise.  Also SYSREM algorithm by  \citealt{2005MNRAS.356.1466T} (implemented in Vartools package, see \citealt{2016A&C....17....1H}) post processing of light curves was applied to remove systemic effects.  

To date, the GPX survey has collected over 750 hours of data from 2016 to mid 2017 and has identified several candidate stars with potential hot Jupiter exoplanets. A follow-up narrow field telescope at Acton Sky Portal is used to determine whether the transits identified are real, and performs a photometry test to determine if the transit depth is achromatic (consistent over different filter bands) using pairs of rotating filters (g' \& i', \textit{$V$} \& \textit{$I_c$}, \textit{$B$} \& \textit{$R_c$}, etc.), similar to methods used by the XO and KELT surveys. Several amateur and professional observers in Europe, USA, and Russia help in verifying transits. Recently, the follow-up network \citep{2017AAS...22914636M} for the KELT exoplanet survey \citep{2012ApJ...761..123S} has started to follow up the brighter GPX candidates. That network, KELT-FUN, obtains photometric confirmation of the shape and ephemeris of the transits to rule out false positives. The candidate GPX stars are currently waiting for spectroscopic follow-up observations with the SOPHIE spectrograph \citep{2009A&A...505..853B} and Tillinghast Reflector Echelle Spectrograph (TRES, \citealt{2007RMxAC..28..129S,2008psa..conf..287F}) via KELT project time to determine whether the achromatic transits are due to hot Jupiters or other companion stars.

We would like to encourage amateur astronomers, who are learning and/or are now very skilled in photometric techniques, to participate in science. One organization, the AAVSO\footnote{\url{www.aavso.org}}, can provide educational materials to train the amateur astronomer interested in using astro CCD imaging to produce scientific data on variable stars and transiting exoplanets. Research tools, described in this article and references herein, provide one means for any engaged person to possibly find transiting exoplanets with the help of some relatively affordable instruments. 

%\section{Conclusions}\label{sec:conclusions}


\acknowledgments
This work was supported in part by the Ministry of Education and Science (the basic part of the State assignment, RK no. AAAA-A17-117030310283-7 and project no. 3.9620.2017/BP), by the Act no. 211 of the Government of the Russian Federation and by the agreement no. 02.A03.21.0006. Sokov E. acknowledges support by the Russian Science Foundation grant No. 14-50-00043 for conducting international photometric observing campaign of the discovered transiting exoplanet candidates. Sokova I. acknowledges support by the Russian Foundation for Basic Research (project No. 17-02-00542). The speckle interferometric observations of the exoplanet candidates with the 6-m telescope of SAO RAS were supported by the Russian Science Foundation grant No. 14-50-00043, area of focus Exoplanets. We thank T\"{U}B\.{I}TAK for a partial support in using T100 telescope with the project number 16AT100-997. The second author (Paul Benni) thanks Bruce Gary, the XO survey, and the KELT survey for furthering his education in exoplanet research. These results use observations collected with the SOPHIE spectrograph on the 1.93-m telescope at Observatoire de Haute-Provence (CNRS), France (program 16A.PNP.HEBR). L.~Delrez acknowledges support from the Gruber Foundation Fellowship. M.~Gillon is F.R.S.-FNRS Research Associate. O.~Demangeon acknowledges the support from Funda\c{c}\~{a}o para a Ci\^encia e a Tecnologia (FCT) through national funds and by FEDER through COMPETE2020 by grants UID/FIS/04434/2013 \& POCI-01-0145-FEDER-007672 and PTDC/FIS-AST/1526/2014 \& POCI-01-0145-FEDER-016886. E.P. acknowledges support from the Research Council of Lithuania (LMT) through grant LAT-08/2016. Authors would also like to express gratitude to Maria Ryavina and the Razvodovs for help with preparation of this paper. Authors thank the anonymous referee for his/her constructive comments, which helped us to improve the quality of the paper.  

This research has made use of the Exoplanet Orbit Database, the Exoplanet Data Explorer at exoplanets.org, Extrasolar Planets Encyclopaedia at exoplanets.eu and the NASA Exoplanet Archive, which is operated by the California Institute of Technology under contract with the National Aeronautics and Space Administration under the Exoplanet Exploration Program. This research made use of Aladin \citep{2000A&AS..143...33B}. IRAF is distributed by the National Optical Astronomy Observatory, which is operated by the Association of Universities for Research in Astronomy, Inc., under cooperative agreement with the National Science Foundation. 


\vspace{5mm}
\bibliography{Burdanov}

\begin{thebibliography}{}
\expandafter\ifx\csname natexlab\endcsname\relax\def\natexlab#1{#1}\fi
\providecommand{\url}[1]{\href{#1}{#1}}

\bibitem[{{Akeson} {et~al.}(2013){Akeson}, {Chen}, {Ciardi}, {Crane}, {Good},
  {Harbut}, {Jackson}, {Kane}, {Laity}, {Leifer}, {Lynn}, {McElroy}, {Papin},
  {Plavchan}, {Ram{\'{\i}}rez}, {Rey}, {von Braun}, {Wittman}, {Abajian},
  {Ali}, {Beichman}, {Beekley}, {Berriman}, {Berukoff}, {Bryden}, {Chan},
  {Groom}, {Lau}, {Payne}, {Regelson}, {Saucedo}, {Schmitz}, {Stauffer},
  {Wyatt}, \& {Zhang}}]{2013PASP..125..989A}
{Akeson}, R.~L., {Chen}, X., {Ciardi}, D., {et~al.} 2013, \pasp, 125, 989

\bibitem[{{Alonso} {et~al.}(2004){Alonso}, {Brown}, {Torres}, {Latham},
  {Sozzetti}, {Mandushev}, {Belmonte}, {Charbonneau}, {Deeg}, {Dunham},
  {O'Donovan}, \& {Stefanik}}]{Alonso2004}
{Alonso}, R., {Brown}, T.~M., {Torres}, G., {et~al.} 2004, \apjl, 613, L153

\bibitem[{{Alsubai} {et~al.}(2013){Alsubai}, {Parley}, {Bramich}, {Horne},
  {Collier Cameron}, {West}, {Sorensen}, {Pollacco}, {Smith}, \&
  {Fors}}]{Alsubai2013}
{Alsubai}, K.~A., {Parley}, N.~R., {Bramich}, D.~M., {et~al.} 2013, \actaa, 63,
  465

\bibitem[{{Auvergne} {et~al.}(2009){Auvergne}, {Bodin}, {Boisnard}, {Buey},
  {Chaintreuil}, {Epstein}, {Jouret}, {Lam-Trong}, {Levacher}, {Magnan},
  {Perez}, {Plasson}, {Plesseria}, {Peter}, {Steller}, {Tiph{\`e}ne}, {Baglin},
  {Agogu{\'e}}, {Appourchaux}, {Barbet}, {Beaufort}, {Bellenger}, {Berlin},
  {Bernardi}, {Blouin}, {Boumier}, {Bonneau}, {Briet}, {Butler}, {Cautain},
  {Chiavassa}, {Costes}, {Cuvilho}, {Cunha-Parro}, {de Oliveira Fialho},
  {Decaudin}, {Defise}, {Djalal}, {Docclo}, {Drummond}, {Dupuis}, {Exil},
  {Faur{\'e}}, {Gaboriaud}, {Gamet}, {Gavalda}, {Grolleau}, {Gueguen},
  {Guivarc'h}, {Guterman}, {Hasiba}, {Huntzinger}, {Hustaix}, {Imbert},
  {Jeanville}, {Johlander}, {Jorda}, {Journoud}, {Karioty}, {Kerjean},
  {Lafond}, {Lapeyrere}, {Landiech}, {Larqu{\'e}}, {Laudet}, {Le Merrer},
  {Leporati}, {Leruyet}, {Levieuge}, {Llebaria}, {Martin}, {Mazy}, {Mesnager},
  {Michel}, {Moalic}, {Monjoin}, {Naudet}, {Neukirchner}, {Nguyen-Kim},
  {Ollivier}, {Orcesi}, {Ottacher}, {Oulali}, {Parisot}, {Perruchot},
  {Piacentino}, {Pinheiro da Silva}, {Platzer}, {Pontet}, {Pradines},
  {Quentin}, {Rohbeck}, {Rolland}, {Rollenhagen}, {Romagnan}, {Russ}, {Samadi},
  {Schmidt}, {Schwartz}, {Sebbag}, {Smit}, {Sunter}, {Tello}, {Toulouse},
  {Ulmer}, {Vandermarcq}, {Vergnault}, {Wallner}, {Waultier}, \&
  {Zanatta}}]{Auvergne2009}
{Auvergne}, M., {Bodin}, P., {Boisnard}, L., {et~al.} 2009, \aap, 506, 411

\bibitem[{{Bakos} {et~al.}(2004){Bakos}, {Noyes}, {Kov{\'a}cs}, {Stanek},
  {Sasselov}, \& {Domsa}}]{Bakos2004}
{Bakos}, G., {Noyes}, R.~W., {Kov{\'a}cs}, G., {et~al.} 2004, \pasp, 116, 266

\bibitem[{{Bakos} {et~al.}(2013){Bakos}, {Csubry}, {Penev}, {Bayliss},
  {Jord{\'a}n}, {Afonso}, {Hartman}, {Henning}, {Kov{\'a}cs}, {Noyes},
  {B{\'e}ky}, {Suc}, {Cs{\'a}k}, {Rabus}, {L{\'a}z{\'a}r}, {Papp}, {S{\'a}ri},
  {Conroy}, {Zhou}, {Sackett}, {Schmidt}, {Mancini}, {Sasselov}, \&
  {Ueltzhoeffer}}]{2013PASP..125..154B}
{Bakos}, G.~{\'A}., {Csubry}, Z., {Penev}, K., {et~al.} 2013, \pasp, 125, 154

\bibitem[{{Baluev} {et~al.}(2015){Baluev}, {Sokov}, {Shaidulin}, {Sokova},
  {Jones}, {Tuomi}, {Anglada-Escud{\'e}}, {Benni}, {Colazo}, {Schneiter},
  {D'Angelo}, {Burdanov}, {Fern{\'a}ndez-Laj{\'u}s}, {Ba{\c s}t{\"u}rk},
  {Hentunen}, \& {Shadick}}]{2015MNRAS.450.3101B}
{Baluev}, R.~V., {Sokov}, E.~N., {Shaidulin}, V.~S., {et~al.} 2015, \mnras,
  450, 3101

\bibitem[{{Baranne} {et~al.}(1996){Baranne}, {Queloz}, {Mayor}, {Adrianzyk},
  {Knispel}, {Kohler}, {Lacroix}, {Meunier}, {Rimbaud}, \&
  {Vin}}]{1996A&AS..119..373B}
{Baranne}, A., {Queloz}, D., {Mayor}, M., {et~al.} 1996, \aaps, 119, 373

\bibitem[{{Boisse} {et~al.}(2010){Boisse}, {Eggenberger}, {Santos}, {Lovis},
  {Bouchy}, {H{\'e}brard}, {Arnold}, {Bonfils}, {Delfosse}, {Desort},
  {D{\'{\i}}az}, {Ehrenreich}, {Forveille}, {Gallenne}, {Lagrange}, {Moutou},
  {Udry}, {Pepe}, {Perrier}, {Perruchot}, {Pont}, {Queloz}, {Santerne},
  {S{\'e}gransan}, \& {Vidal-Madjar}}]{2010A&A...523A..88B}
{Boisse}, I., {Eggenberger}, A., {Santos}, N.~C., {et~al.} 2010, \aap, 523, A88

\bibitem[{{Bonnarel} {et~al.}(2000){Bonnarel}, {Fernique}, {Bienaym{\'e}},
  {Egret}, {Genova}, {Louys}, {Ochsenbein}, {Wenger}, \&
  {Bartlett}}]{2000A&AS..143...33B}
{Bonnarel}, F., {Fernique}, P., {Bienaym{\'e}}, O., {et~al.} 2000, \aaps, 143,
  33

\bibitem[{{Borucki} {et~al.}(2010){Borucki}, {Koch}, {Basri}, {Batalha},
  {Brown}, {Caldwell}, {Caldwell}, {Christensen-Dalsgaard}, {Cochran},
  {DeVore}, {Dunham}, {Dupree}, {Gautier}, {Geary}, {Gilliland}, {Gould},
  {Howell}, {Jenkins}, {Kondo}, {Latham}, {Marcy}, {Meibom}, {Kjeldsen},
  {Lissauer}, {Monet}, {Morrison}, {Sasselov}, {Tarter}, {Boss}, {Brownlee},
  {Owen}, {Buzasi}, {Charbonneau}, {Doyle}, {Fortney}, {Ford}, {Holman},
  {Seager}, {Steffen}, {Welsh}, {Rowe}, {Anderson}, {Buchhave}, {Ciardi},
  {Walkowicz}, {Sherry}, {Horch}, {Isaacson}, {Everett}, {Fischer}, {Torres},
  {Johnson}, {Endl}, {MacQueen}, {Bryson}, {Dotson}, {Haas}, {Kolodziejczak},
  {Van Cleve}, {Chandrasekaran}, {Twicken}, {Quintana}, {Clarke}, {Allen},
  {Li}, {Wu}, {Tenenbaum}, {Verner}, {Bruhweiler}, {Barnes}, \&
  {Prsa}}]{Borucki2010}
{Borucki}, W.~J., {Koch}, D., {Basri}, G., {et~al.} 2010, Science, 327, 977

\bibitem[{{Bouchy} {et~al.}(2009){Bouchy}, {H{\'e}brard}, {Udry}, {Delfosse},
  {Boisse}, {Desort}, {Bonfils}, {Eggenberger}, {Ehrenreich}, {Forveille},
  {Lagrange}, {Le Coroller}, {Lovis}, {Moutou}, {Pepe}, {Perrier}, {Pont},
  {Queloz}, {Santos}, {S{\'e}gransan}, \& {Vidal-Madjar}}]{2009A&A...505..853B}
{Bouchy}, F., {H{\'e}brard}, G., {Udry}, S., {et~al.} 2009, \aap, 505, 853

\bibitem[{{Burdanov} {et~al.}(2014){Burdanov}, {Krushinsky}, \&
  {Popov}}]{2014AstBu..69..368B}
{Burdanov}, A.~Y., {Krushinsky}, V.~V., \& {Popov}, A.~A. 2014, Astrophysical
  Bulletin, 69, 368

\bibitem[{{Burdanov} {et~al.}(2016){Burdanov}, {Benni}, {Krushinsky}, {Popov},
  {Sokov}, {Sokova}, {Rusov}, {Lyashenko}, {Ivanov}, {Moiseev}, {Rastegaev},
  {Dyachenko}, {Balega}, {Ba{\c s}t{\"u}rk}, {{\"O}zavc{\i}}, {Puchalski},
  {Marchini}, {Naves}, {Shadick}, \& {Bretton}}]{2016MNRAS.461.3854B}
{Burdanov}, A.~Y., {Benni}, P., {Krushinsky}, V.~V., {et~al.} 2016, \mnras,
  461, 3854

\bibitem[{{Burrows}(2014)}]{2014Natur.513..345B}
{Burrows}, A.~S. 2014, \nat, 513, 345

\bibitem[{{Chang} {et~al.}(2010){Chang}, {Gu}, \&
  {Bodenheimer}}]{2010ApJ...708.1692C}
{Chang}, S.-H., {Gu}, P.-G., \& {Bodenheimer}, P.~H. 2010, \apj, 708, 1692

\bibitem[{{Correia} \& {Laskar}(2010)}]{2010exop.book..239C}
{Correia}, A.~C.~M., \& {Laskar}, J. 2010, {Tidal Evolution of Exoplanets}, ed.
  S.~{Seager}, 239--266

\bibitem[{{Croll}(2012)}]{2012JAVSO..40..456C}
{Croll}, B. 2012, Journal of the American Association of Variable Star
  Observers (JAAVSO), 40, 456

\bibitem[{{Crouzet}(2017)}]{2017arXiv171207275C}
{Crouzet}, N. 2017, ArXiv e-prints, arXiv:1712.07275

\bibitem[{{Cutri} {et~al.}(2003){Cutri}, {Skrutskie}, {van Dyk}, {Beichman},
  {Carpenter}, {Chester}, {Cambresy}, {Evans}, {Fowler}, {Gizis}, {Howard},
  {Huchra}, {Jarrett}, {Kopan}, {Kirkpatrick}, {Light}, {Marsh}, {McCallon},
  {Schneider}, {Stiening}, {Sykes}, {Weinberg}, {Wheaton}, {Wheelock}, \&
  {Zacarias}}]{2003yCat.2246....0C}
{Cutri}, R.~M., {Skrutskie}, M.~F., {van Dyk}, S., {et~al.} 2003, VizieR Online
  Data Catalog, 2246

\bibitem[{{Eastman} {et~al.}(2010){Eastman}, {Siverd}, \&
  {Gaudi}}]{2010PASP..122..935E}
{Eastman}, J., {Siverd}, R., \& {Gaudi}, B.~S. 2010, \pasp, 122, 935

\bibitem[{{Enoch} {et~al.}(2010){Enoch}, {Collier Cameron}, {Parley}, \&
  {Hebb}}]{2010A&A...516A..33E}
{Enoch}, B., {Collier Cameron}, A., {Parley}, N.~R., \& {Hebb}, L. 2010, \aap,
  516, A33

\bibitem[{{F{\H u}r{\'e}sz} {et~al.}(2008){F{\H u}r{\'e}sz}, {Szentgyorgyi}, \&
  {Meibom}}]{2008psa..conf..287F}
{F{\H u}r{\'e}sz}, G., {Szentgyorgyi}, A.~H., \& {Meibom}, S. 2008, in
  Precision Spectroscopy in Astrophysics, ed. N.~C. {Santos}, L.~{Pasquini},
  A.~C.~M. {Correia}, \& M.~{Romaniello}, 287--290

\bibitem[{{Fischer} {et~al.}(2012){Fischer}, {Schwamb}, {Schawinski},
  {Lintott}, {Brewer}, {Giguere}, {Lynn}, {Parrish}, {Sartori}, {Simpson},
  {Smith}, {Spronck}, {Batalha}, {Rowe}, {Jenkins}, {Bryson}, {Prsa},
  {Tenenbaum}, {Crepp}, {Morton}, {Howard}, {Beleu}, {Kaplan}, {Vannispen},
  {Sharzer}, {Defouw}, {Hajduk}, {Neal}, {Nemec}, {Schuepbach}, \&
  {Zimmermann}}]{2012MNRAS.419.2900F}
{Fischer}, D.~A., {Schwamb}, M.~E., {Schawinski}, K., {et~al.} 2012, \mnras,
  419, 2900

\bibitem[{{Fortney} {et~al.}(2007){Fortney}, {Marley}, \&
  {Barnes}}]{2007ApJ...659.1661F}
{Fortney}, J.~J., {Marley}, M.~S., \& {Barnes}, J.~W. 2007, \apj, 659, 1661

\bibitem[{{Gelman} \& {Rubin}(1992)}]{1992StaSc...7..457G}
{Gelman}, A., \& {Rubin}, D.~B. 1992, Statistical Science, 7, 457

\bibitem[{{Gillon} {et~al.}(2011){Gillon}, {Doyle}, {Lendl}, {Maxted},
  {Triaud}, {Anderson}, {Barros}, {Bento}, {Collier-Cameron}, {Enoch}, {Faedi},
  {Hellier}, {Jehin}, {Magain}, {Montalb{\'a}n}, {Pepe}, {Pollacco}, {Queloz},
  {Smalley}, {Segransan}, {Smith}, {Southworth}, {Udry}, {West}, \&
  {Wheatley}}]{2011A&A...533A..88G}
{Gillon}, M., {Doyle}, A.~P., {Lendl}, M., {et~al.} 2011, \aap, 533, A88

\bibitem[{{Gillon} {et~al.}(2012){Gillon}, {Triaud}, {Fortney}, {Demory},
  {Jehin}, {Lendl}, {Magain}, {Kabath}, {Queloz}, {Alonso}, {Anderson},
  {Collier Cameron}, {Fumel}, {Hebb}, {Hellier}, {Lanotte}, {Maxted},
  {Mowlavi}, \& {Smalley}}]{2012A&A...542A...4G}
{Gillon}, M., {Triaud}, A.~H.~M.~J., {Fortney}, J.~J., {et~al.} 2012, \aap,
  542, A4

\bibitem[{{Gorbovskoy} {et~al.}(2013){Gorbovskoy}, {Lipunov}, {Kornilov},
  {Belinski}, {Kuvshinov}, {Tyurina}, {Sankovich}, {Krylov}, {Shatskiy},
  {Balanutsa}, {Chazov}, {Kuznetsov}, {Zimnukhov}, {Shumkov}, {Shurpakov},
  {Senik}, {Gareeva}, {Pruzhinskaya}, {Tlatov}, {Parkhomenko}, {Dormidontov},
  {Krushinsky}, {Punanova}, {Zalozhnyh}, {Popov}, {Burdanov}, {Yazev},
  {Budnev}, {Ivanov}, {Konstantinov}, {Gress}, {Chuvalaev}, {Yurkov},
  {Sergienko}, {Kudelina}, {Sinyakov}, {Karachentsev}, {Moiseev}, \&
  {Fatkhullin}}]{Gorbovskoy2013}
{Gorbovskoy}, E.~S., {Lipunov}, V.~M., {Kornilov}, V.~G., {et~al.} 2013,
  Astronomy Reports, 57, 233

\bibitem[{{Guyon} {et~al.}(2014){Guyon}, {Walawender}, {Jovanovic},
  {Butterfield}, {Gee}, \& {Mery}}]{2014SPIE.9145E..3VG}
{Guyon}, O., {Walawender}, J., {Jovanovic}, N., {et~al.} 2014, in \procspie,
  Vol. 9145, Ground-based and Airborne Telescopes V, 91453V

\bibitem[{{Han} {et~al.}(2014){Han}, {Wang}, {Wright}, {Feng}, {Zhao},
  {Fakhouri}, {Brown}, \& {Hancock}}]{2014PASP..126..827H}
{Han}, E., {Wang}, S.~X., {Wright}, J.~T., {et~al.} 2014, \pasp, 126, 827

\bibitem[{{Hartman} \& {Bakos}(2016)}]{2016A&C....17....1H}
{Hartman}, J.~D., \& {Bakos}, G.~{\'A}. 2016, Astronomy and Computing, 17, 1

\bibitem[{{H{\'e}brard} {et~al.}(2008){H{\'e}brard}, {Bouchy}, {Pont},
  {Loeillet}, {Rabus}, {Bonfils}, {Moutou}, {Boisse}, {Delfosse}, {Desort},
  {Eggenberger}, {Ehrenreich}, {Forveille}, {Lagrange}, {Lovis}, {Mayor},
  {Pepe}, {Perrier}, {Queloz}, {Santos}, {S{\'e}gransan}, {Udry}, \&
  {Vidal-Madjar}}]{2008A&A...488..763H}
{H{\'e}brard}, G., {Bouchy}, F., {Pont}, F., {et~al.} 2008, \aap, 488, 763

\bibitem[{{H{\'e}brard} {et~al.}(2011){H{\'e}brard}, {Ehrenreich}, {Bouchy},
  {Delfosse}, {Moutou}, {Arnold}, {Boisse}, {Bonfils}, {D{\'{\i}}az},
  {Eggenberger}, {Forveille}, {Lagrange}, {Lovis}, {Pepe}, {Perrier}, {Queloz},
  {Santerne}, {Santos}, {S{\'e}gransan}, {Udry}, \&
  {Vidal-Madjar}}]{2011A&A...527L..11H}
{H{\'e}brard}, G., {Ehrenreich}, D., {Bouchy}, F., {et~al.} 2011, \aap, 527,
  L11

\bibitem[{{Henden} {et~al.}(2016){Henden}, {Templeton}, {Terrell}, {Smith},
  {Levine}, \& {Welch}}]{2016yCat.2336....0H}
{Henden}, A.~A., {Templeton}, M., {Terrell}, D., {et~al.} 2016, VizieR Online
  Data Catalog, 2336

\bibitem[{{H{\o}g} {et~al.}(2000){H{\o}g}, {Fabricius}, {Makarov}, {Urban},
  {Corbin}, {Wycoff}, {Bastian}, {Schwekendiek}, \&
  {Wicenec}}]{2000A&A...355L..27H}
{H{\o}g}, E., {Fabricius}, C., {Makarov}, V.~V., {et~al.} 2000, \aap, 355, L27

\bibitem[{{Holman} {et~al.}(2006){Holman}, {Winn}, {Latham}, {O'Donovan},
  {Charbonneau}, {Bakos}, {Esquerdo}, {Hergenrother}, {Everett}, \&
  {P{\'a}l}}]{2006ApJ...652.1715H}
{Holman}, M.~J., {Winn}, J.~N., {Latham}, D.~W., {et~al.} 2006, \apj, 652, 1715

\bibitem[{{Howell} {et~al.}(2014){Howell}, {Sobeck}, {Haas}, {Still},
  {Barclay}, {Mullally}, {Troeltzsch}, {Aigrain}, {Bryson}, {Caldwell},
  {Chaplin}, {Cochran}, {Huber}, {Marcy}, {Miglio}, {Najita}, {Smith},
  {Twicken}, \& {Fortney}}]{2014PASP..126..398H}
{Howell}, S.~B., {Sobeck}, C., {Haas}, M., {et~al.} 2014, \pasp, 126, 398

\bibitem[{{Hroch}(2014)}]{2014ascl.soft02006H}
{Hroch}, F. 2014, {Munipack: General astronomical image processing software},
  Astrophysics Source Code Library, , , ascl:1402.006

\bibitem[{{Kov{\'a}cs} {et~al.}(2002){Kov{\'a}cs}, {Zucker}, \&
  {Mazeh}}]{2002A&A...391..369K}
{Kov{\'a}cs}, G., {Zucker}, S., \& {Mazeh}, T. 2002, \aap, 391, 369

\bibitem[{{Madhusudhan} {et~al.}(2016){Madhusudhan}, {Ag{\'u}ndez}, {Moses}, \&
  {Hu}}]{2016SSRv..205..285M}
{Madhusudhan}, N., {Ag{\'u}ndez}, M., {Moses}, J.~I., \& {Hu}, Y. 2016, \ssr,
  205, 285

\bibitem[{{Maksimov} {et~al.}(2009){Maksimov}, {Balega}, {Dyachenko},
  {Malogolovets}, {Rastegaev}, \& {Semernikov}}]{2009AstBu..64..296M}
{Maksimov}, A.~F., {Balega}, Y.~Y., {Dyachenko}, V.~V., {et~al.} 2009,
  Astrophysical Bulletin, 64, 296

\bibitem[{{Mandel} \& {Agol}(2002)}]{2002ApJ...580L.171M}
{Mandel}, K., \& {Agol}, E. 2002, \apjl, 580, L171

\bibitem[{{McCullough} {et~al.}(2005){McCullough}, {Stys}, {Valenti},
  {Fleming}, {Janes}, \& {Heasley}}]{McCullough2005}
{McCullough}, P.~R., {Stys}, J.~E., {Valenti}, J.~A., {et~al.} 2005, \pasp,
  117, 783

\bibitem[{{McLeod} {et~al.}(2017){McLeod}, {Melton}, {Stassun}, \& {KELT
  Collaboration}}]{2017AAS...22914636M}
{McLeod}, K.~K., {Melton}, C., {Stassun}, K.~G., \& {KELT Collaboration}. 2017,
  in American Astronomical Society Meeting Abstracts, Vol. 229, American
  Astronomical Society Meeting Abstracts, 146.36

\bibitem[{{Moutou} {et~al.}(2011){Moutou}, {D{\'{\i}}az}, {Udry},
  {H{\'e}brard}, {Bouchy}, {Santerne}, {Ehrenreich}, {Arnold}, {Boisse},
  {Bonfils}, {Delfosse}, {Eggenberger}, {Forveille}, {Lagrange}, {Lovis},
  {Martinez}, {Pepe}, {Perrier}, {Queloz}, {Santos}, {S{\'e}gransan},
  {Toublanc}, {Troncin}, {Vanhuysse}, \& {Vidal-Madjar}}]{2011A&A...533A.113M}
{Moutou}, C., {D{\'{\i}}az}, R.~F., {Udry}, S., {et~al.} 2011, \aap, 533, A113

\bibitem[{{Murray} \& {Correia}(2010)}]{2010exop.book...15M}
{Murray}, C.~D., \& {Correia}, A.~C.~M. 2010, {Keplerian Orbits and Dynamics of
  Exoplanets}, ed. S.~{Seager}, 15--23

\bibitem[{{Pepe} {et~al.}(2002){Pepe}, {Mayor}, {Galland}, {Naef}, {Queloz},
  {Santos}, {Udry}, \& {Burnet}}]{2002A&A...388..632P}
{Pepe}, F., {Mayor}, M., {Galland}, F., {et~al.} 2002, \aap, 388, 632

\bibitem[{{Pollacco} {et~al.}(2008){Pollacco}, {Skillen}, {Collier Cameron},
  {Loeillet}, {Stempels}, {Bouchy}, {Gibson}, {Hebb}, {H{\'e}brard}, {Joshi},
  {McDonald}, {Smalley}, {Smith}, {Street}, {Udry}, {West}, {Wilson},
  {Wheatley}, {Aigrain}, {Alsubai}, {Benn}, {Bruce}, {Christian}, {Clarkson},
  {Enoch}, {Evans}, {Fitzsimmons}, {Haswell}, {Hellier}, {Hickey}, {Hodgkin},
  {Horne}, {Hrudkov{\'a}}, {Irwin}, {Kane}, {Keenan}, {Lister}, {Maxted},
  {Mayor}, {Moutou}, {Norton}, {Osborne}, {Parley}, {Pont}, {Queloz}, {Ryans},
  \& {Simpson}}]{2008MNRAS.385.1576P}
{Pollacco}, D., {Skillen}, I., {Collier Cameron}, A., {et~al.} 2008, \mnras,
  385, 1576

\bibitem[{{Pollacco} {et~al.}(2006){Pollacco}, {Skillen}, {Collier Cameron},
  {Christian}, {Hellier}, {Irwin}, {Lister}, {Street}, {West}, {Anderson},
  {Clarkson}, {Deeg}, {Enoch}, {Evans}, {Fitzsimmons}, {Haswell}, {Hodgkin},
  {Horne}, {Kane}, {Keenan}, {Maxted}, {Norton}, {Osborne}, {Parley}, {Ryans},
  {Smalley}, {Wheatley}, \& {Wilson}}]{2006PASP..118.1407P}
{Pollacco}, D.~L., {Skillen}, I., {Collier Cameron}, A., {et~al.} 2006, \pasp,
  118, 1407

\bibitem[{{Popov} {et~al.}(2015){Popov}, {Burdanov}, {Zubareva}, {Krushinsky},
  {Avvakumova}, \& {Ivanov}}]{2015PZP....15....7P}
{Popov}, A.~A., {Burdanov}, A.~Y., {Zubareva}, A.~M., {et~al.} 2015, Peremennye
  Zvezdy Prilozhenie, 15

\bibitem[{{Rauer} {et~al.}(2014){Rauer}, {Catala}, {Aerts}, {Appourchaux},
  {Benz}, {Brandeker}, {Christensen-Dalsgaard}, {Deleuil}, {Gizon}, {Goupil},
  {G{\"u}del}, {Janot-Pacheco}, {Mas-Hesse}, {Pagano}, {Piotto}, {Pollacco},
  {Santos}, {Smith}, {Su{\'a}rez}, {Szab{\'o}}, {Udry}, {Adibekyan}, {Alibert},
  {Almenara}, {Amaro-Seoane}, {Eiff}, {Asplund}, {Antonello}, {Barnes},
  {Baudin}, {Belkacem}, {Bergemann}, {Bihain}, {Birch}, {Bonfils}, {Boisse},
  {Bonomo}, {Borsa}, {Brand{\~a}o}, {Brocato}, {Brun}, {Burleigh}, {Burston},
  {Cabrera}, {Cassisi}, {Chaplin}, {Charpinet}, {Chiappini}, {Church},
  {Csizmadia}, {Cunha}, {Damasso}, {Davies}, {Deeg}, {D{\'{\i}}az}, {Dreizler},
  {Dreyer}, {Eggenberger}, {Ehrenreich}, {Eigm{\"u}ller}, {Erikson}, {Farmer},
  {Feltzing}, {de Oliveira Fialho}, {Figueira}, {Forveille}, {Fridlund},
  {Garc{\'{\i}}a}, {Giommi}, {Giuffrida}, {Godolt}, {Gomes da Silva},
  {Granzer}, {Grenfell}, {Grotsch-Noels}, {G{\"u}nther}, {Haswell}, {Hatzes},
  {H{\'e}brard}, {Hekker}, {Helled}, {Heng}, {Jenkins}, {Johansen},
  {Khodachenko}, {Kislyakova}, {Kley}, {Kolb}, {Krivova}, {Kupka}, {Lammer},
  {Lanza}, {Lebreton}, {Magrin}, {Marcos-Arenal}, {Marrese}, {Marques},
  {Martins}, {Mathis}, {Mathur}, {Messina}, {Miglio}, {Montalban}, {Montalto},
  {Monteiro}, {Moradi}, {Moravveji}, {Mordasini}, {Morel}, {Mortier},
  {Nascimbeni}, {Nelson}, {Nielsen}, {Noack}, {Norton}, {Ofir}, {Oshagh},
  {Ouazzani}, {P{\'a}pics}, {Parro}, {Petit}, {Plez}, {Poretti}, {Quirrenbach},
  {Ragazzoni}, {Raimondo}, {Rainer}, {Reese}, {Redmer}, {Reffert},
  {Rojas-Ayala}, {Roxburgh}, {Salmon}, {Santerne}, {Schneider}, {Schou},
  {Schuh}, {Schunker}, {Silva-Valio}, {Silvotti}, {Skillen}, {Snellen}, {Sohl},
  {Sousa}, {Sozzetti}, {Stello}, {Strassmeier}, {{\v S}vanda}, {Szab{\'o}},
  {Tkachenko}, {Valencia}, {Van Grootel}, {Vauclair}, {Ventura}, {Wagner},
  {Walton}, {Weingrill}, {Werner}, {Wheatley}, \&
  {Zwintz}}]{2014ExA....38..249R}
{Rauer}, H., {Catala}, C., {Aerts}, C., {et~al.} 2014, Experimental Astronomy,
  38, 249

\bibitem[{{Ricker} {et~al.}(2015){Ricker}, {Winn}, {Vanderspek}, {Latham},
  {Bakos}, {Bean}, {Berta-Thompson}, {Brown}, {Buchhave}, {Butler}, {Butler},
  {Chaplin}, {Charbonneau}, {Christensen-Dalsgaard}, {Clampin}, {Deming},
  {Doty}, {De Lee}, {Dressing}, {Dunham}, {Endl}, {Fressin}, {Ge}, {Henning},
  {Holman}, {Howard}, {Ida}, {Jenkins}, {Jernigan}, {Johnson}, {Kaltenegger},
  {Kawai}, {Kjeldsen}, {Laughlin}, {Levine}, {Lin}, {Lissauer}, {MacQueen},
  {Marcy}, {McCullough}, {Morton}, {Narita}, {Paegert}, {Palle}, {Pepe},
  {Pepper}, {Quirrenbach}, {Rinehart}, {Sasselov}, {Sato}, {Seager},
  {Sozzetti}, {Stassun}, {Sullivan}, {Szentgyorgyi}, {Torres}, {Udry}, \&
  {Villasenor}}]{2015JATIS...1a4003R}
{Ricker}, G.~R., {Winn}, J.~N., {Vanderspek}, R., {et~al.} 2015, Journal of
  Astronomical Telescopes, Instruments, and Systems, 1, 014003

\bibitem[{{Sanchis-Ojeda} \& {Winn}(2011)}]{2011ApJ...743...61S}
{Sanchis-Ojeda}, R., \& {Winn}, J.~N. 2011, \apj, 743, 61

\bibitem[{{Sanchis-Ojeda} {et~al.}(2011){Sanchis-Ojeda}, {Winn}, {Holman},
  {Carter}, {Osip}, \& {Fuentes}}]{2011ApJ...733..127S}
{Sanchis-Ojeda}, R., {Winn}, J.~N., {Holman}, M.~J., {et~al.} 2011, \apj, 733,
  127

\bibitem[{{Santerne} {et~al.}(2014){Santerne}, {H{\'e}brard}, {Deleuil},
  {Havel}, {Correia}, {Almenara}, {Alonso}, {Arnold}, {Barros}, {Behrend},
  {Bernasconi}, {Boisse}, {Bonomo}, {Bouchy}, {Bruno}, {Damiani},
  {D{\'{\i}}az}, {Gravallon}, {Guillot}, {Labrevoir}, {Montagnier}, {Moutou},
  {Rinner}, {Santos}, {Abe}, {Audejean}, {Bendjoya}, {Gillier}, {Gregorio},
  {Martinez}, {Michelet}, {Montaigut}, {Poncy}, {Rivet}, {Rousseau}, {Roy},
  {Suarez}, {Vanhuysse}, \& {Verilhac}}]{2014A&A...571A..37S}
{Santerne}, A., {H{\'e}brard}, G., {Deleuil}, M., {et~al.} 2014, \aap, 571, A37

\bibitem[{{Schneider} {et~al.}(2011){Schneider}, {Dedieu}, {Le Sidaner},
  {Savalle}, \& {Zolotukhin}}]{2011A&A...532A..79S}
{Schneider}, J., {Dedieu}, C., {Le Sidaner}, P., {Savalle}, R., \&
  {Zolotukhin}, I. 2011, \aap, 532, A79

\bibitem[{{Schwarz}(1978)}]{1978AnSta...6..461S}
{Schwarz}, G. 1978, Annals of Statistics, 6, 461

\bibitem[{{Seager} \& {Mall{\'e}n-Ornelas}(2003)}]{2003ApJ...585.1038S}
{Seager}, S., \& {Mall{\'e}n-Ornelas}, G. 2003, \apj, 585, 1038

\bibitem[{{Siverd} {et~al.}(2012){Siverd}, {Beatty}, {Pepper}, {Eastman},
  {Collins}, {Bieryla}, {Latham}, {Buchhave}, {Jensen}, {Crepp}, {Street},
  {Stassun}, {Gaudi}, {Berlind}, {Calkins}, {DePoy}, {Esquerdo}, {Fulton},
  {F{\H u}r{\'e}sz}, {Geary}, {Gould}, {Hebb}, {Kielkopf}, {Marshall}, {Pogge},
  {Stanek}, {Stefanik}, {Szentgyorgyi}, {Trueblood}, {Trueblood}, {Stutz}, \&
  {van Saders}}]{2012ApJ...761..123S}
{Siverd}, R.~J., {Beatty}, T.~G., {Pepper}, J., {et~al.} 2012, \apj, 761, 123

\bibitem[{{Southworth}(2011)}]{2011MNRAS.417.2166S}
{Southworth}, J. 2011, \mnras, 417, 2166

\bibitem[{{Szentgyorgyi} \& {Fur{\'e}sz}(2007)}]{2007RMxAC..28..129S}
{Szentgyorgyi}, A.~H., \& {Fur{\'e}sz}, G. 2007, in Revista Mexicana de
  Astronomia y Astrofisica Conference Series, Vol.~28, Revista Mexicana de
  Astronomia y Astrofisica Conference Series, ed. S.~{Kurtz}, 129--133

\bibitem[{{Tamuz} {et~al.}(2005){Tamuz}, {Mazeh}, \&
  {Zucker}}]{2005MNRAS.356.1466T}
{Tamuz}, O., {Mazeh}, T., \& {Zucker}, S. 2005, \mnras, 356, 1466

\bibitem[{{Tody}(1986)}]{Tody1986}
{Tody}, D. 1986, in Proceedings of the Meeting, Tucson, AZ, March 4-8, 1986,
  Vol. 627, Instrumentation in astronomy VI, ed. D.~L. {Crawford} (Bellingham,
  WA: Society of Photo-Optical Instrumentation Engineers (SPIE) Conference
  Series), 733

\bibitem[{{Triaud} {et~al.}(2010){Triaud}, {Collier Cameron}, {Queloz},
  {Anderson}, {Gillon}, {Hebb}, {Hellier}, {Loeillet}, {Maxted}, {Mayor},
  {Pepe}, {Pollacco}, {S{\'e}gransan}, {Smalley}, {Udry}, {West}, \&
  {Wheatley}}]{2010A&A...524A..25T}
{Triaud}, A.~H.~M.~J., {Collier Cameron}, A., {Queloz}, D., {et~al.} 2010,
  \aap, 524, A25

\bibitem[{{Valenti} \& {Piskunov}(1996)}]{1996A&AS..118..595V}
{Valenti}, J.~A., \& {Piskunov}, N. 1996, \aaps, 118, 595

\bibitem[{{Wheatley} {et~al.}(2017){Wheatley}, {West}, {Goad}, {Jenkins},
  {Pollacco}, {Queloz}, {Rauer}, {Udry}, {Watson}, {Chazelas}, {Eigmuller},
  {Lambert}, {Genolet}, {McCormac}, {Walker}, {Armstrong}, {Bayliss}, {Bento},
  {Bouchy}, {Burleigh}, {Cabrera}, {Casewell}, {Chaushev}, {Chote},
  {Csizmadia}, {Erikson}, {Faedi}, {Foxell}, {Gansicke}, {Gillen}, {Grange},
  {Gunther}, {Hodgkin}, {Jackman}, {Jordan}, {Louden}, {Metrailler}, {Moyano},
  {Nielsen}, {Osborn}, {Poppenhaeger}, {Raddi}, {Raynard}, {Smith}, {Soto}, \&
  {Titz-Weider}}]{2017arXiv171011100W}
{Wheatley}, P.~J., {West}, R.~G., {Goad}, M.~R., {et~al.} 2017, ArXiv e-prints,
  arXiv:1710.11100

\bibitem[{{Winn}(2010)}]{Winn2010}
{Winn}, J.~N. 2010, ArXiv e-prints, arXiv:1001.2010

\bibitem[{{Wright} {et~al.}(2012){Wright}, {Marcy}, {Howard}, {Johnson},
  {Morton}, \& {Fischer}}]{2012ApJ...753..160W}
{Wright}, J.~T., {Marcy}, G.~W., {Howard}, A.~W., {et~al.} 2012, \apj, 753, 160

\end{thebibliography}
%\end{thebibliography}



\end{document}